\documentclass[trackchanges]{aastex701}

\usepackage{amsmath}

\newcommand{\Ro}{\mathrm{Ro}}
\newcommand{\Roc}{\mathrm{Ro_C}}

\begin{document}

\title{Influence of penetration depth on jets on giant planets: equatorial jet direction, jet numbers, and jet energy fraction}

\author[orcid=0000-0002-2624-8579]{Yaoxuan Zeng}
\affiliation{Department of the Geophysical Sciences, The University of Chicago, Chicago, 60637, USA.}
\email[show]{yxzeng@uchicago.edu}  

\author[orcid=0000-0002-4615-3702]{Wanying Kang} 
\affiliation{Earth, Atmospheric and Planetary Science Department, Massachusetts Institute of Technology, Cambridge, 02139, USA.}
\email{wanying@mit.edu}

\author[orcid=0000-0003-3589-5249]{Glenn R. Flierl}
\affiliation{Earth, Atmospheric and Planetary Science Department, Massachusetts Institute of Technology, Cambridge, 02139, USA.}
\email{glenn@lake.mit.edu}

\author[orcid=0000-0002-5971-8995]{Geoffrey K. Vallis}
\affiliation{Mathematics and Statistics, University of Exeter, Exeter, UK}
\email{G.Vallis@exeter.ac.uk}

\begin{abstract}

It remains puzzling why, despite their similar nature, Jupiter and Saturn possess a prograde equatorial jet, whereas Uranus and Neptune have a retrograde one. To understand this discrepancy, we use a two-dimensional quasi-geostrophic model to explore how the jet penetration depth, regulated by Ohmic dissipation, influences the structure and organization of jet patterns. When jets penetrate deeply into the planetary interior, the effective planetary vorticity gradient $\beta$ becomes negative near the equator and decreases equatorward due to spherical geometry. This $\beta$ profile favors dynamical modes that transport eastward momentum toward the equator, producing a prograde equatorial jet, as observed on Jupiter and Saturn. In contrast, relatively shallow systems favor a retrograde equatorial jet. In our simulations, the equatorial jet direction is primarily controlled by the gradient of $\beta$, as predicted by Stochastic Structural Stability Theory, rather than by its sign, as suggested by Potential Vorticity mixing. If this mechanism applies to Uranus and Neptune, the observed jet structure may suggest the presence of a stratified or Ohmic dissipation layer near their surfaces. At mid-latitudes, jet widths are constrained by the Rhines scale, yielding a scaling that explains the presence of multiple jets on Jupiter and Saturn and a single jet per hemisphere on Uranus and Neptune. Lastly, we examine how planetary parameters influence the partitioning of energy between jets and eddies. Stronger energy input, faster rotation, smaller planetary radius, or weaker large-scale damping lead to a larger fraction of the total energy in zonal jets, resulting in smoother jet structures.

\end{abstract}

\keywords{Jet formation --- Giant planets --- Quasi-geostrophic}

\section{Introduction} 

\setcounter{footnote}{0}

\begin{figure*}[b!]
\centering
\includegraphics[width=.9\linewidth]{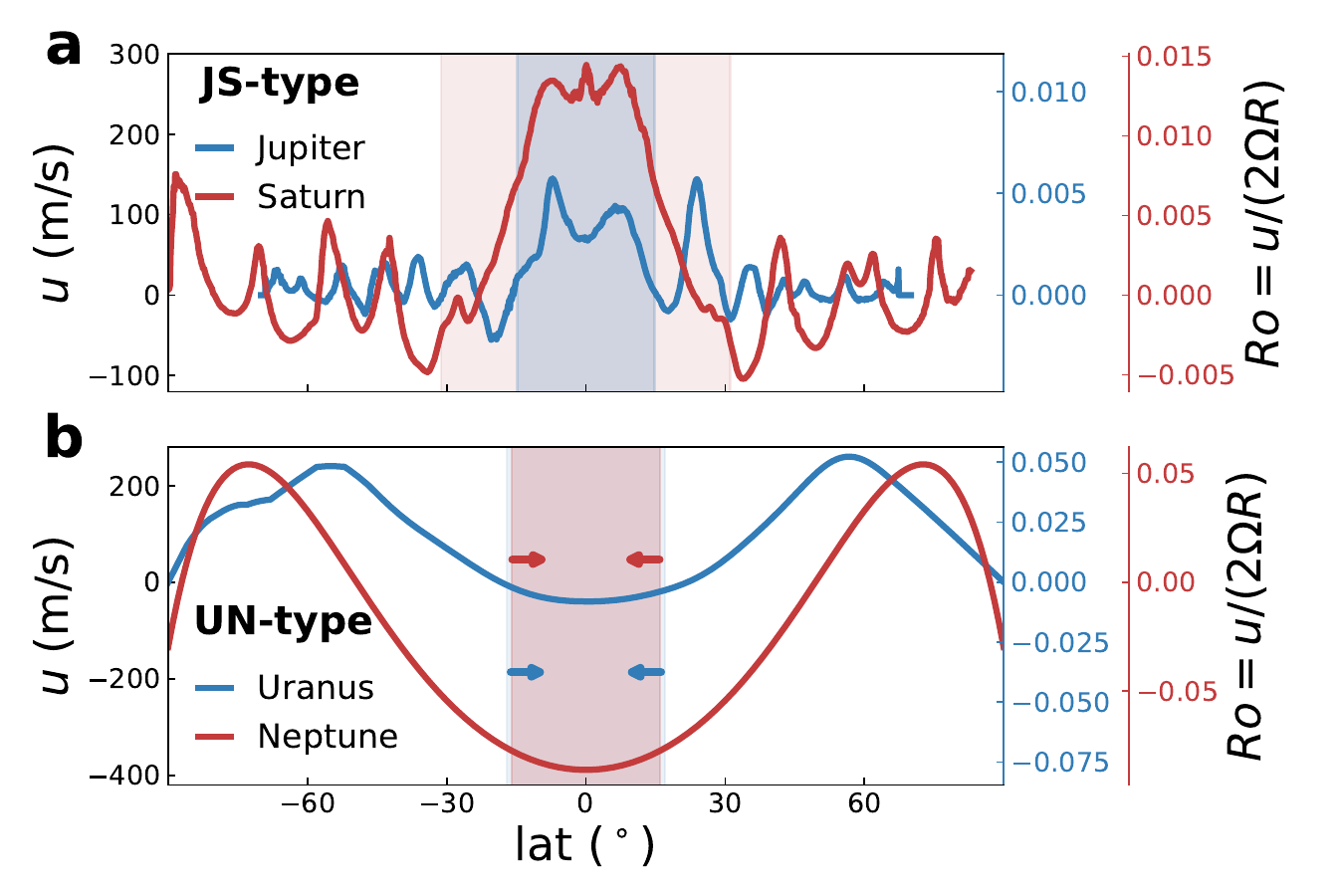}
\caption{\textbf{Observed surface zonal wind profile on giant planets.} The JS-type jets are shown in \textbf{a} and the UN-type jets are shown in \textbf{b}. The shading regions indicate regions outside the TC, with blue representing Jupiter in \textbf{a} and Uranus in \textbf{b}, and red representing Saturn in \textbf{a} and Neptune in \textbf{b}. For Uranus and Neptune, the TC latitude is less well constrained, and we adopt the upper-limit estimate from \citet{kaspi2013atmospheric}. The Rossby number of the flow, $\Ro=U/(2\Omega R)$ where $U$ is the velocity, $\Omega$ is the rotation rate, and $R$ is the planetary radius, is also shown. For Jupiter, data were obtained from Juno's third perijove pass on December 11, 2016 \citep{tollefson2017changes}. For Saturn, measurements were taken from Cassini ISS CB2 and CB3 images spanning 2004 to 2008 \citep{garcia2011saturn}, corrected with the rotation rate in \cite{helled2015saturn}. For Uranus and Neptune, due to limited observations, fitted curves are used, based on \cite{sromovsky2015high} and \cite{sromovsky1993dynamics}, respectively.}
\label{fig:u-observation}
\end{figure*}

Zonal jets are expected to form on rapidly rotating giant planets, and their existence on Jupiter, Saturn, Uranus and Neptune has been captured by various missions during flybys, orbiters, and space based telescopes \citep{limaye1986jupiter,sromovsky1993dynamics,sanchez2000saturn,garcia2001study,garcia2011saturn,sromovsky2015high,tollefson2017changes,sanchez2019zonal}. It is revealed that the jet profiles of these planets fall into two distinct categories. Jupiter and Saturn exhibit multiple jets in each hemisphere, with particularly strong prograde jets near the equator (hereafter, JS-type jet, Fig.~\ref{fig:u-observation}a). In contrast, Uranus and Neptune have only one jet in each hemisphere, and their equatorial jets are retrograde (hereafter, UN-type jet, Fig.~\ref{fig:u-observation}b). The differences in the four giant planets' jet characteristics clearly pose an intriguing puzzle about jet formation mechanisms, and addressing this puzzle may offer us an opportunity to infer the interior properties of these planets.

Several mechanisms have been proposed to explain their formation. The first considers an atmosphere energized by convection due to intrinsic heating \citep{busse1970thermal,busse1976simple,ingersoll1982motion,aurnou2001strong,christensen2001zonal,christensen2002zonal,yano2005deep,heimpel2005simulation,heimpel2007turbulent,aurnou2007effects,kaspi2009deep,jones2009compressible,soderlund2013turbulent,gastine2013zonal,gastine2013solar,gastine2014zonal,yadav2020deep,yadav2022global,camisassa2022solar,duer2023gas,duer2024depth,duer2025gas}. In this framework, the direction of the equatorial jet is found to be controlled by the convective Rossby number, $\Roc$, which measures the relative importance of rotation versus buoyancy forcing \citep{aurnou2007effects}. At low $\Roc$ (i.e., under rapid rotation), convection organizes into columnar rolls aligned with the rotation axis. These structures, known as Busse modes \citep{busse1970thermal,busse1976simple}, dominate the equatorial regions outside the tangent cylinder (TC; see Fig.~\ref{fig:sketch-beta-h}). They tilt in the prograde direction with height under the influence of the spherical geometry outside the TC \citep{busse1982differential,busse1983convection}, pumping prograde momentum to sustain a prograde jet at the equatorial surface. When buoyancy force is strong enough to break the constraint by planetary rotation ($\Roc>1$), Busse modes lose coherence. Turbulent mixing then transports low-angular-momentum gas from the deep interior toward the surface, giving rise to a retrograde equatorial jet instead \citep{aurnou2007effects,gastine2013zonal,gastine2013solar,camisassa2022solar}.

The second mechanism considers jets driven by the momentum transport of Rossby waves \citep{williams1978planetary,cho1996morphogenesis,williams2003barotropic,williams2003jet,showman2007numerical,lian2008deep,scott2008equatorial,schneider2009formation,liu2010mechanisms,lian2010generation,liu2011convective,warneford2017super,young2019simulating,spiga2020global}. In this framework, the jet direction depends on the sign of $\beta$ and the location of the wave source. For Rossby waves, the eddy momentum flux, $\overline{u'v'}$, satisfies $\overline{u'v'} \propto -c_{g_y}/\beta$, where $u'$ and $v'$ are anomalies of zonal and meridional velocities, and $c_{g_y}$ is the meridional group velocity representing the direction of wave energy propagation. Wave energy always propagates away from its source region. Therefore, for $\beta>0$, the momentum flux has the opposite sign of the group velocity and converges toward the wave source, leading to the formation of a prograde jet at the forcing latitude and retrograde jets where the energy is dissipated; for $\beta<0$, the entire jet pattern reverses sign \citep{vallis2017atmospheric}. In the conventional case of $\beta>0$, which has been the focus of previous studies, the jet structure further depends on the location of the wave source: if wave generation is dominated by equatorial forcing (e.g., convection), the resulting momentum convergence produces a prograde equatorial jet, whereas if mid-latitude forcing by baroclinic instability dominates, momentum converges at mid-latitudes, yielding prograde jets there and a retrograde jet at the equator \citep{schneider2009formation,liu2010mechanisms,liu2011convective}.

The third mechanism interprets jet formation as a consequence of Potential Vorticity (PV) mixing \citep{yano2005deep,vallis2019essentials}. In this framework, PV, as a conserved quantity, is homogenized by turbulence. The planetary PV gradient is balanced by the PV gradient induced by the flow, giving rise to parabolic zonal jets, $u\sim \beta(y-y_c)^2/2$, where $y_c$ is the jet center\footnote{This jet shape is derived for infinite deformation radius ($L_d \to \infty$) with a constant $\beta$ in \cite{yano2005deep} and \cite{vallis2019essentials}. For finite $L_d$, the jet becomes hyperbolic \citep{marcus1994jupiter}, but the qualitative behavior remains the same.}. Assuming hemispheric symmetry about the equator and that the jet vanishes at latitude $y=y_j$, the jet profile becomes $u = \beta (y^2-y_j^2)/2$. In this case, when $\beta > 0$, $u(y=0)<0$, and a retrograde equatorial jet forms, whereas when $\beta<0$, a prograde jet forms at the equator. The equatorial jet direction therefore depends only on the sign of $\beta$, regardless of the forcing latitude.

The fourth mechanism of jet formation arises from transient mode growth driven by stochastic excitation, as described by the Stochastic Structural Stability Theory \citep[SSST;][]{farrell2003structural,farrell2007structure,farrell2008formation,farrell2009stochastic,srinivasan2012zonostrophic,constantinou2014emergence}. In SSST, imposed stochastic forcing excites dynamical modes that undergo transient growth followed by decay. These modes are tilted by the planetary $\beta$ gradient, and their statistically averaged contribution produces a nonzero eddy momentum flux that leads to jet formation. In contrast to the Rossby wave propagation and PV mixing mechanisms described above, the jet direction predicted by SSST depends on the gradient of $\beta$ (i.e., whether $\beta$ increases or decreases toward the equator) rather than on its absolute sign. To some extent, this dependence on the $\beta$ gradient is analogous to the Busse mode mechanism, in which the tilting direction of convective columns depends on whether the bounding surface is convex or concave, leading to different signs of the effective $\beta$ gradient \citep{busse1982differential,busse1983convection,duer2025gas}.

The $\beta$ profile plays a critical role in all aforementioned mechanisms. In rapidly rotating systems representative of giant planet atmospheres, characterized by small convective Rossby numbers (Table~\ref{tab:planet_parameter}), the flow follows the Taylor–Proudman theorem, with motions largely invariant along the rotation axis and organized into Taylor columns. As fluid moves from the rotation axis toward the equatorial surface, the height of Taylor columns is constrained by the boundaries and therefore changes with latitude, modifying the PV and setting the effective planetary PV gradient $\beta$. The columns transition from stretching to compression at the TC, producing a sign change of $\beta$ (Fig.~\ref{fig:sketch-beta-h}). The latitude of the TC is set by the jet penetration depth, which is in turn governed by the planet’s interior structure \citep[e.g.,][]{fortney2010interior,helled2017internal,morf2025icy}, as electrical conductivity increases with depth and eventually becomes sufficiently high for Ohmic dissipation to damp the flow \citep{liu2006interaction}. On Jupiter and Saturn, gravitational field measurements suggest jet penetration depths of approximately 2400~km \citep{kaspi2018jupiter,duer2020range,galanti2021combined,kaspi2023observational,cao2023strong} and 8700~km \citep{galanti2019saturn,galanti2021combined}, respectively, consistent with estimates based on conductivity profiles \citep{liu2006interaction,french2012ab,kaspi2020comparison}. For Uranus and Neptune, the jet depths are far less well constrained: gravity data are limited, leading to an upper-limit estimation of 1100~km for both planets \citep{kaspi2013atmospheric}; their conductivity profiles are also poorly constrained, with maximum estimated penetration depths of 2600~km for Uranus and 2000~km for Neptune \citep{soyuer2020constraining}. The differences in the jet penetration depth, and hence in the $\beta$ profile, may help explain the differences between JS-type and UN-type jets.

The goal of this work is two-fold. First, we examine how the jet penetration depth affects jet formation on giant planets. Second, we assess whether the direction of the equatorial jet is primarily controlled by the absolute sign of $\beta$, as suggested by the PV mixing mechanism, or by the gradient of $\beta$, as predicted by SSST. To address these questions, we employ a 1.5-layer Quasi-Geostrophic Potential Vorticity (QGPV) model to study jet formation on giant planets. This idealized model allows us to prescribe the effect of domain depth, which determines the $\beta$ profile, and to encompass multiple mechanisms within a unified framework. The model is designed to capture how energy input at small scales self-organizes into coherent eddy structures that redistribute momentum and sustain zonal jets. Rather than explicitly resolving energy generation processes, i.e., deep convection or baroclinic instability, we directly inject energy as spatially homogeneous stochastic forcing \citep{smith2004local}. This setup allows the model to capture PV mixing and SSST mechanisms. In principle, it can also represent Rossby wave propagation, although the uniform forcing does not distinguish the dominant forcing latitude, which is essential for determining the jet profile in this mechanism. Our model does not capture Busse modes, as these require convective dynamics to be explicitly represented. One may consider applying spatially varying forcing to examine the Rossby wave propagation mechanism, or inject tilted eddies as energy input to mimic the Busse mode mechanism, but these are beyond the scope of this study. Section~\ref{sec:method} describes the model setup. In Section~\ref{sec:direction}, we demonstrate how the jet penetration depth controls the direction of the equatorial jet. Sections~\ref{sec:number} and \ref{sec:smoothness} discuss the number and smoothness of jets, respectively. Section~\ref{sec:discussion} provides discussion and concluding remarks.

\section{Method}\label{sec:method}

\begin{figure*}[b!]
\centering
\includegraphics[width=0.95\textwidth]{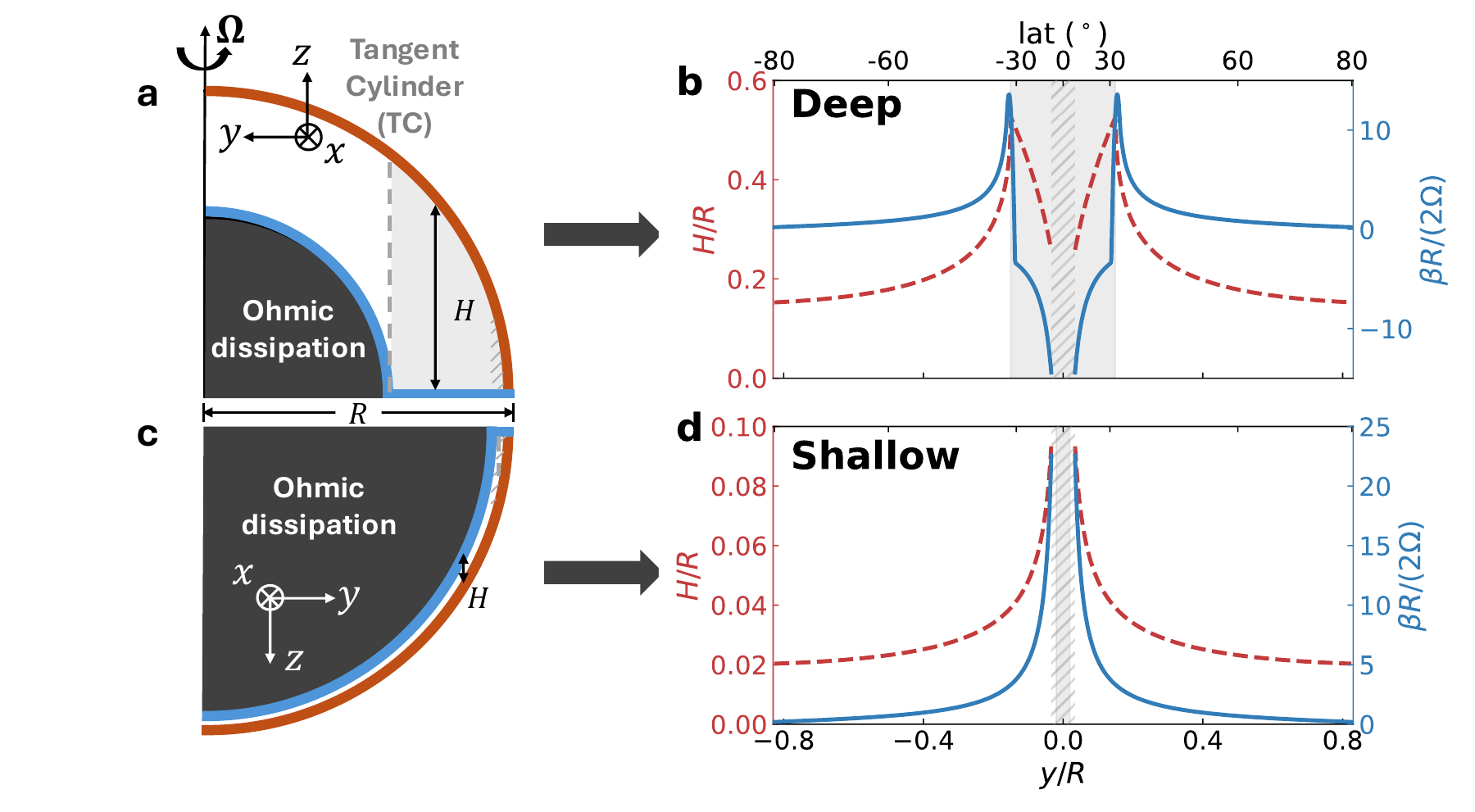}
\caption{\textbf{Geometry of the atmosphere on giant planets and the corresponding $\beta$ profile.} The upper row shows the deep scenario ($D=0.15R$, \textbf{a} and \textbf{b}), and the lower row shows the shallow scenario ($D=0.02R$, \textbf{c} and \textbf{d}), where $D$ is the jet penetration depth and $R$ is the planetary radius. In the schematics (\textbf{a} and \textbf{c}), the black shaded region indicates areas where jets are damped by Ohmic dissipation. The column depth $H$ is defined as the distance from the surface (brown line) to the bottom, marked by the Ohmic dissipation level inside the TC and a virtual equatorial plane outside the TC (blue line). The gray dashed line indicates the TC, where the cylindrical air column contacts the Ohmic dissipation layer. The light gray shading indicates regions outside the TC. The gray hatched region marks the area excluded from our simulation domain to avoid the singularity at the equator. Inside the TC, $H$ decreases with latitude and $\beta$ is positive, whereas outside the TC, $H$ increases with latitude and $\beta$ becomes negative. All values in \textbf{b} and \textbf{d} are non-dimensionalized using planetary radius and rotation (see APPENDIX~\ref{app:model} for detailed derivation).}
\label{fig:sketch-beta-h}
\end{figure*}

All four giant planets are in the rapidly rotating limit (RRL),  characterized by convective Rossby numbers much smaller than unity \citep{aurnou2020connections} (Table~\ref{tab:planet_parameter}). In this regime, motions are largely invariant along the rotation axis, following the Taylor–Proudman theorem \citep[e.g.,][]{heimpel2005simulation,vallis2019essentials,aurnou2020connections}. This allows us to reduce the axial dimension of the system by integrating the governing equations along the rotation axis. We assume the atmosphere is isentropic, reflecting the strong convective mixing that efficiently homogenizes entropy, and barotropic along the rotation axis, consistent with the Taylor–Proudman theorem in the rapidly rotating limit. Under these assumptions, the governing dynamics reduce to a 2-D QGPV equation (a detailed derivation is provided in APPENDIX~\ref{app:model}):

\begin{equation}\label{eq:governing}
    \frac{Dq }{D t} = \mathcal{F}(\varepsilon) - \mu \zeta - \nu_h \nabla^4 \zeta, \ \ q = \zeta - \frac{\psi}{L_d^2} + \int_0^y \beta dy',
\end{equation}

\begin{deluxetable*}{ccccc}
\tablenum{1}
\tablecaption{Planetary parameters of the four giant planets in the solar system. \label{tab:planet_parameter}}
\tablewidth{0pt}
\tablehead{
\colhead{Parameters} & \colhead{Jupiter} & \colhead{Saturn} & \colhead{Uranus} & \colhead{Neptune}
}
\startdata
$R$ ($10^3$ km) & $71$ & $60$ & $26$ & $25$ \\
$D$ ($10^3$ km) & $2.38 \pm 0.14$ & $8.74 \pm 0.10$ & $<1.10$ & $<1.10$ \\
$Q_{t}$ (W~m$^{-2}$) & $7.49 \pm 0.16$ & $2.84 \pm 0.20$ & $0.042 \pm 0.047$ & $0.433 \pm 0.046$ \\
$T_{t}$ (K) & $103.9 \pm 0.6$ & $77.4 \pm 1.6$ & $58.2 \pm 1.0$ & $46.6 \pm 1.1$ \\
$g$ (m~s$^{-2}$) & $22.88$ & $9.5$ & $8.69$ & $11$ \\
$\Omega$ (10$^{-4}$ s$^{-1}$) & $1.774$ & $1.647$ & $0.975$ & $0.909$ \\
$R_d$ (J~kg$^{-1}$~K$^{-1}$) & $3605.4$ & $4016.4$ & $3149.2$ & $3197.7$ \\
$\varepsilon$ (10$^{-8}$ m$^2$~s$^{-3}$) & $142.0 \pm 6.9$ & $5.1 \pm 0.8$ & $>3.4 \pm 4.0$ & $>15.8 \pm 3.3$ \\
$\varepsilon/(R^{2} (2\Omega)^{3})$ (10$^{-13}$) & $62.6 \pm 3.1$ & $3.9 \pm 0.6$ & $>70.3 \pm 83.9$ & $>428.5 \pm 89.2$ \\
$\mathrm{Ro_{C}} \sim (\varepsilon (2\Omega)^{-3}D^{-2})^{1/5} $ & 0.02 & 0.007 & $>0.02$ & $>0.03$ \\
\hline
$L_e$ (km) & $20.6 \pm 0.2$ & $18.2 \pm 0.7$ & $>8.6 \pm 2.6$ & $>13.4 \pm 0.7$ \\
$L_\beta$ ($10^3$ km) & $1.071 \pm 0.010$ & $0.519 \pm 0.017$ & $>0.392 \pm 0.094$ & $>0.547 \pm 0.023$ \\
$L_{Rh}$ ($10^3$ km) & $7.32$ & $9.87$ & $16.7$ & $12.1$ \\
$L_{Rh,eq}$ ($10^3$ km) & $5.25$ & $8.66$ & $8.46$ & $6.11$ \\
\enddata
\tablecomments{The radius ($R$), gravity ($g$), and rotation rate ($\Omega$) data are from \cite{sanchez2018atmospheric}. The gas constant ($R_d$) values are from \cite{lodders1998planetary}. The intrinsic heat flux ($Q_t$) and equivalent temperature ($T_{t}$) data are from \cite{li2018less} for Jupiter, \cite{wang2024cassini} for Saturn, and \cite{pearl1991albedo} for Uranus and Neptune. Jet penetration depth ($D$) estimates are taken from \cite{kaspi2018jupiter} for Jupiter, \cite{galanti2019saturn} for Saturn, and \cite{kaspi2013atmospheric} for Uranus and Neptune. Detailed derivations of the energy injection rate $\varepsilon$ and characteristic length scales are provided in APPENDIX~\ref{app:epsilon}.}
\end{deluxetable*}


\noindent where $q$ is the potential vorticity (PV), $\psi$ is the stream function satisfying $u=-\partial \psi/\partial y$ and $v=\partial \psi/\partial x$, $\zeta \equiv \partial v/\partial x - \partial u/\partial y$ is the vorticity, and $u$ and $v$ are velocities along $x$ and $y$, respectively. $\Omega$ is the planetary rotation rate, and $\beta=(-2\Omega/H)(dH/dy)$ is the planetary PV gradient due to the spherical geometry, determined by $H$, the column depth parallel to the rotation axis (Fig.~\ref{fig:sketch-beta-h}; see APPENDIX~\ref{app:model} for detailed derivation). $L_d$ is the Rossby deformation radius. $\mathcal{F}$ is the stochastic eddy forcing term used to parameterize the energy injection by convection or baroclinic eddies, whose magnitude depends on the energy injection rate $\varepsilon$. The forcing wavelength is set as small as possible ($4 L_{grid}$, where $L_{grid}$ is the grid size) to allow adequate resolution of the energy cascade to larger scales \citep{maltrud1991energy,vallis1993generation}. $\mu$ is the linear damping coefficient representing jet dissipation by the planet’s electromagnetic field at the bottom of the domain \citep[Ohmic dissipation;][]{liu2008constraints,perna2010magnetic, french2012ab, soyuer2020constraining}. The strong linear damping is only present inside the TC, where the atmospheric column is in direct contact with the Ohmic dissipation layer. $\nu_h$ is the hyperviscosity coefficient to ensure numerical stability. A detailed description of the model configuration can be found in APPENDIX~\ref{app:model}. 

We numerically integrate Equation~\ref{eq:governing} using Dedalus, which solves initial-value partial differential equations using spectral methods \citep{burns2020dedalus}. The equation is non-dimensionalized with a length scale of the planetary radius $R$ and a time scale based on planetary rotation, $(2\Omega)^{-1}$. To avoid the singularity in the $\beta$ profile at the equator, the domain is set between $15^\circ$ and $80^\circ$, excluding the outer $0.034R$ in the $y$ direction. A Cartesian coordinate system and double-periodic boundary conditions are applied, allowing for Fast Fourier Transform (FFT), which significantly improves numerical efficiency. We define the positive $y$ direction as increasing with latitude from the South Pole ($-90^\circ$) to the North Pole (90$^\circ$), while maintaining the same positive $x$ direction in both hemispheres. As a result, the positive $z$ direction changes sign between the two hemispheres (Fig.~\ref{fig:sketch-beta-h}). To resolve the inverse cascade from small-scale eddies to jets, we use a resolution of 512 grid points in the $x$ direction and 1024 grid points in the $y$ direction, with the same grid spacing in two directions, $L_{grid}=1.155 \times 10^{-3}$.

We test two different jet penetration depths: $D=0.02R$ for the shallow case and $D=0.15R$ for the deep case. The critical latitude of the TC, $\phi_c$, where the cylindrical atmospheric column touches the Ohmic dissipation level, is $11.5^\circ$ for the shallow case and $31.8^\circ$ for the deep case. It is numerically difficult to resolve the outside TC region in the shallow case with the resolution applied in this study, so this region is excluded from the simulations. Consequently, in the shallow case, the domain contains no region outside the TC where $|\phi| < \phi_c$. The atmospheric columns outside the TC in the northern and southern hemispheres are connected, implying a symmetry in the flow. Therefore, we enforce symmetry\footnote{$u(-y)=u(y)$ is symmetric, $v(-y)=-v(y)$ and $\zeta(-y)=-\zeta(y)$ are anti-symmetric across the equator outside the TC.} of the flow outside the TC after each timestep. We explore a range of parameter regimes using four different energy injection rates: $\varepsilon = 5 \times 10^{-10}$, $2.9 \times 10^{-11}$, $1.7 \times 10^{-12}$, and $10^{-13}$, representative of estimates for the four giant planets (Table~\ref{tab:planet_parameter}; see APPENDIX~\ref{app:epsilon} for details). We also consider two different linear damping coefficients, $\mu = 5 \times 10^{-8}$ and $10^{-7}$, such that the estimated jet speed, $(\varepsilon / \mu)^{1/2}$, spans the range of observed values (Fig.~\ref{fig:u-observation}). Two different deformation radii are considered, one with $L_d\rightarrow\infty$ representing a bounded unstratified atmosphere, and another with a finite $L_d = 0.16 L_{Rh}$, representing Jupiter-like stratification (see APPENDIX~\ref{app:model} for details).

Equilibrating simulations with such low energy injection rates and damping rates is numerically challenging. To address this, we first run simulations with both $\varepsilon$ and $\mu$ increased by a factor of 1000, ensuring that the jet speed magnitude remains unchanged, to reach an equilibrium state. We then continue the simulations with the smaller $\varepsilon$ and $\mu$. Despite this approach, the adjustment of the flow remains computationally expensive, so simulations are run until the zonal-mean zonal velocity exhibits no significant changes (details can be found in APPENDIX~\ref{app:model}). Unless otherwise noted, the results presented in the main text are long-term averages to reduce high-frequency variability.

\section{Direction of the equatorial jet}\label{sec:direction}


\begin{figure*}[t]
\centering
\includegraphics[width=1.0\textwidth]{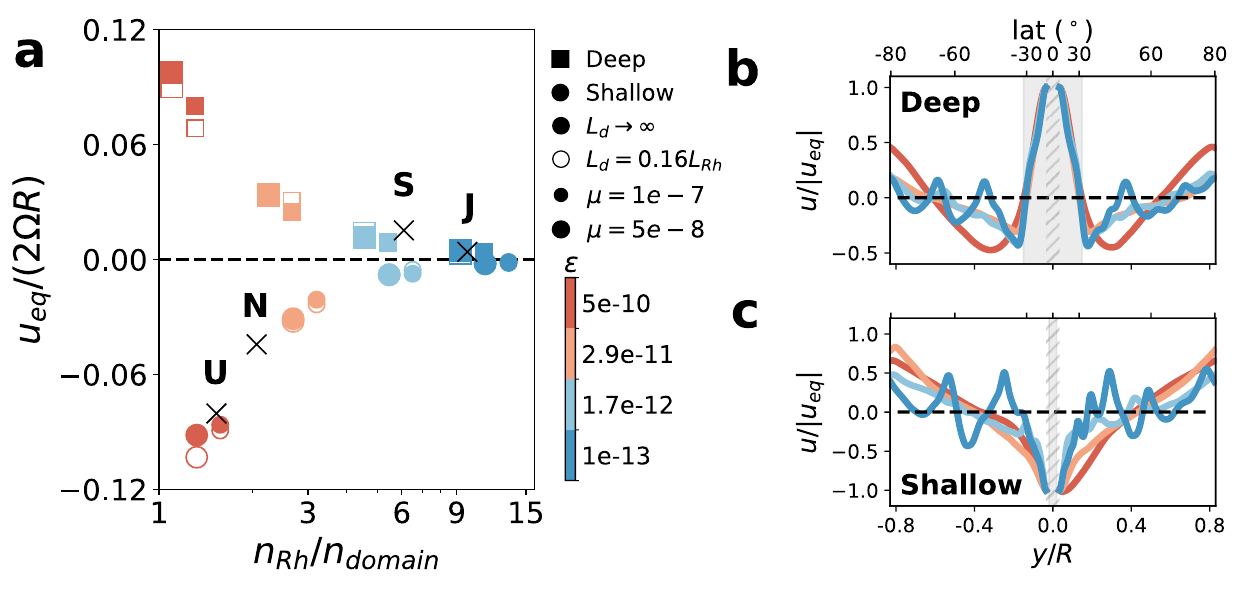}
\caption{\textbf{The direction of the equatorial jet.} \textbf{a}, Non-dimensionalized equatorial zonal jet speed ($u_{eq}/(2\Omega R)$) as a function of the ratio of the Rhines wavenumber $n_{Rh}$ to the domain wavenumber $n_{domain}$. Since our model domain does not extend exactly to the equator, we take the velocity at the lowest simulated latitude to represent the equatorial jet speed. \textbf{b} and \textbf{c}, Zonal-mean zonal velocity profiles, normalized by the equatorial jet speed magnitude ($u / |u_{eq}|$), with a deep (\textbf{b}) or shallow (\textbf{c}) $\beta$ profile. Parameters for Jupiter (J), Saturn (S), Uranus (U), and Neptune (N) are marked with crosses. In \textbf{b} and \textbf{c}, the gray shading indicates regions outside the TC, and the gray hatched region marks the area excluded from our simulation domain. The $\beta$ profiles applied for the deep and shallow cases are shown as blue lines in Fig.~\ref{fig:sketch-beta-h}b and~d, respectively. In \textbf{b} and \textbf{c}, only $\mu = 10^{-7}$ simulations are shown. The line colors correspond to marker colors in \textbf{a}, representing different energy injection rates.}\label{fig:direction}
\end{figure*}

\begin{figure*}[t]
\centering
\includegraphics[width=1.0\textwidth]{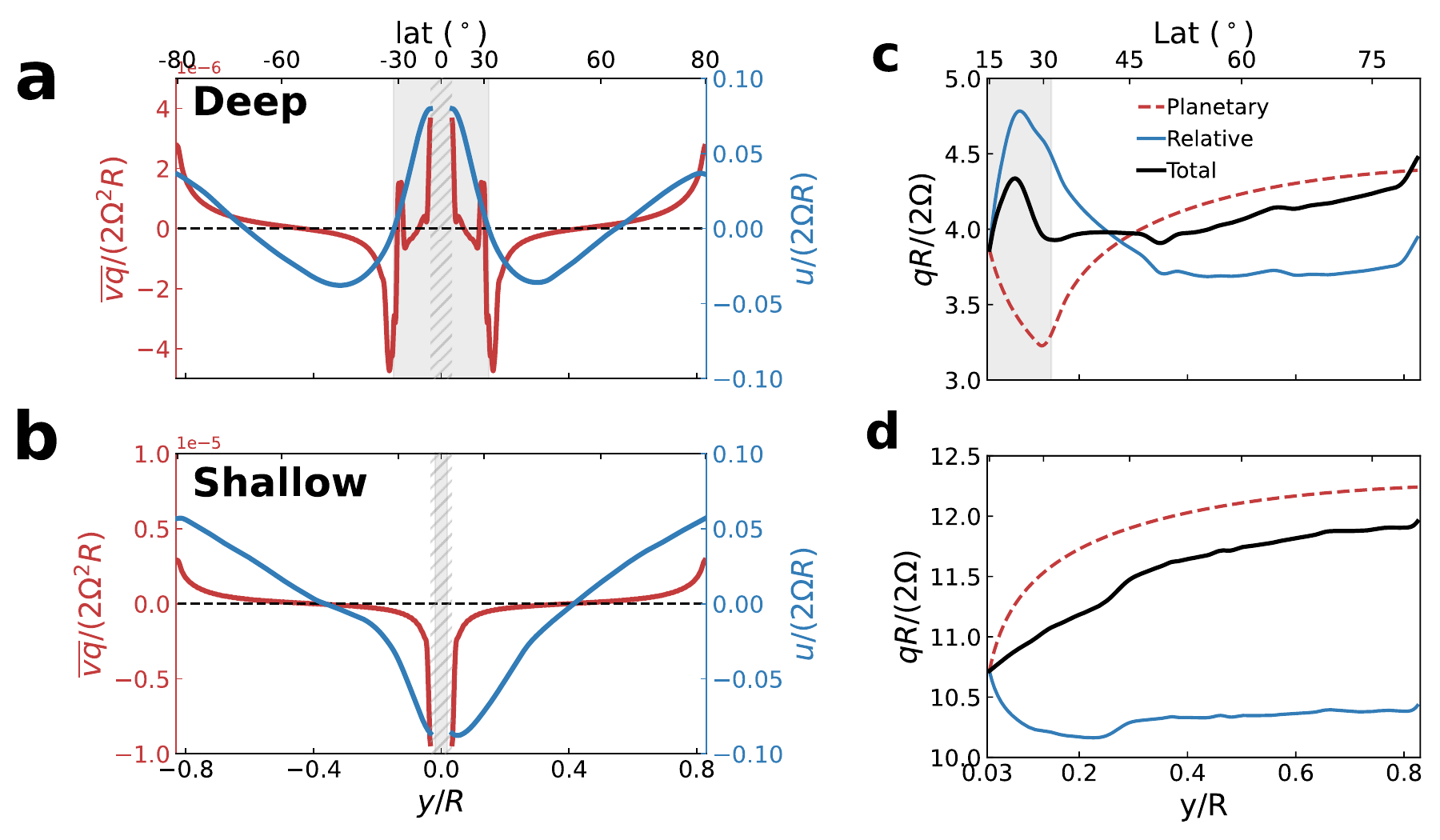}
\caption{\textbf{Examination of different jet formation mechanisms.} \textbf{a} and \textbf{b}, zonal-mean non-dimensionlized eddy momentum flux convergence ($\overline{vq}/(2\Omega^2R)$, red lines) calculated from the deep and shallow $\beta$ profiles and zonal jet speed ($u/(2\Omega R)$, blue lines) profiles as a function of latitude in the simulation with $L_d \to \infty$, $\varepsilon=5\times10^{-10}$, and $\alpha=10^{-7}$. \textbf{c} and \textbf{d}, non-dimensionalized zonal-mean PV profile ($qR/(2\Omega)$) in the northern hemisphere for the same simulations in \textbf{a} and \textbf{b}, respectively, where red line denotes planetary PV ($\overline{q_p}=\int{\beta}dy=-2\Omega \ln{H}$), blue line denotes jet-induced relative PV ($\overline{q_r}=-\partial \overline{u}/\partial y-\overline{\psi}/L_d^2$), and black line denotes total PV ($\overline{q}=\overline{q_p}+\overline{q_r}$), where overline denotes zonal average. A constant offset (equal to $\overline{q}_p$ at the equator) is added to $\overline{q}_r$ for better comparison.}\label{fig:Mechanism_compare}
\end{figure*}

\begin{figure*}[t]
\centering
\includegraphics[width=0.75\textwidth]{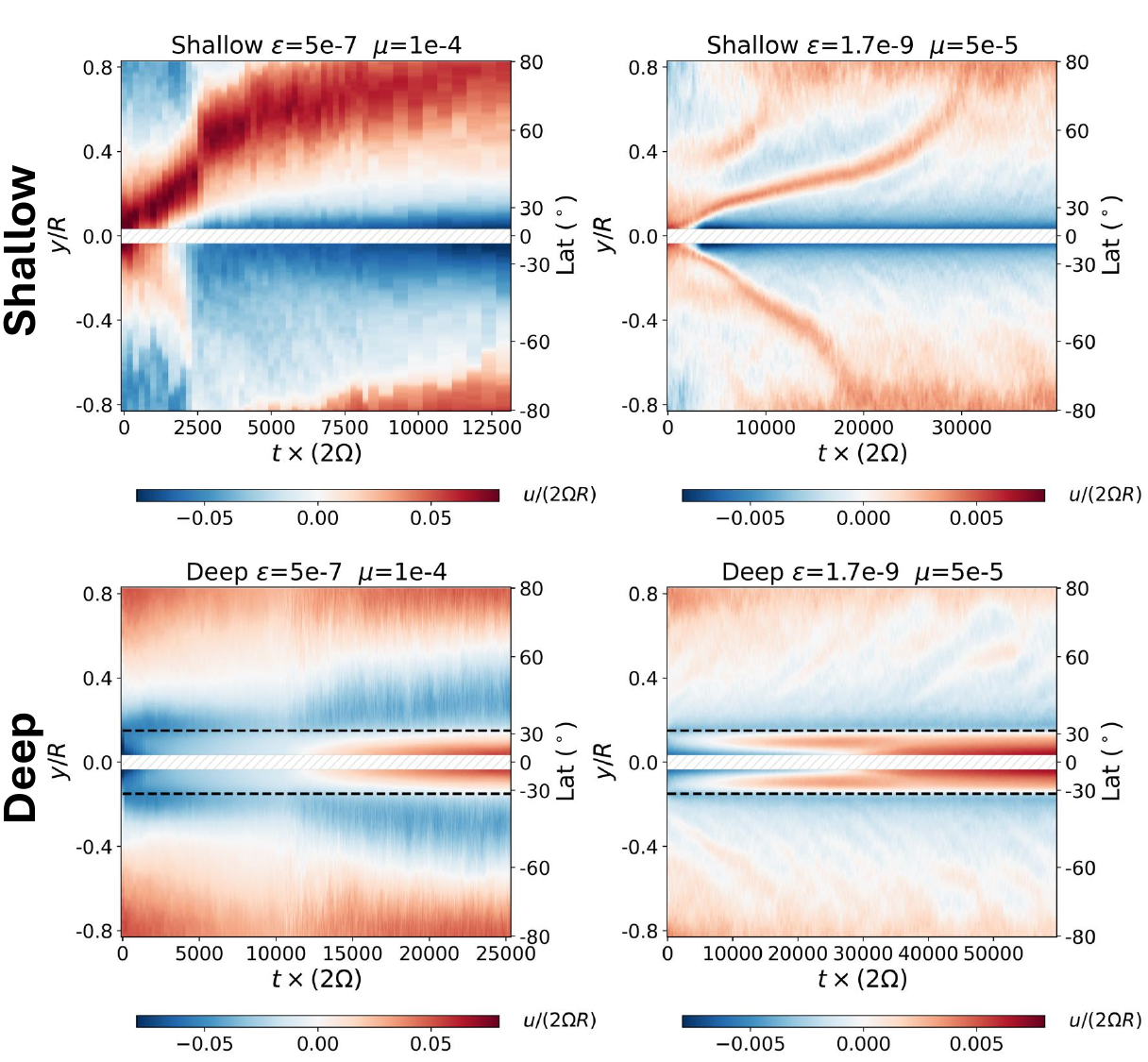}
\caption{\textbf{Sensitivity to initial conditions.} Evolution of the zonal-mean zonal velocity for simulations initialized with reversed jets (see text for details). The horizontal axis represents simulation time, and the vertical axis represents latitude. The hatched region near the equator marks the latitudes that are not included in the model domain, and the black dashed lines indicate the TC latitude in the deep cases.}
\label{figS2:IC}
\end{figure*}

\begin{figure*}[b]
\centering
\includegraphics[width=1.0\textwidth]{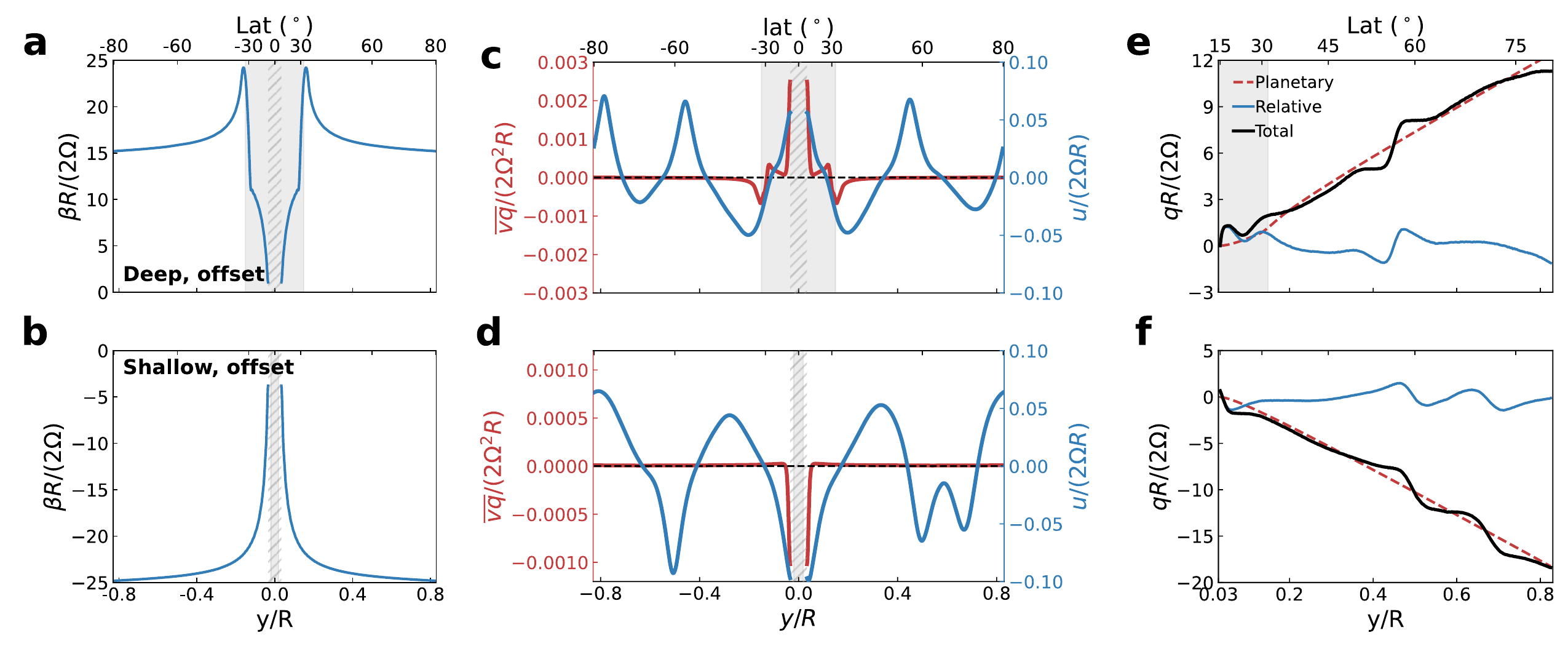}
\caption{\textbf{Sensitivity tests on the sign and gradient of $\beta$.} \textbf{a} and \textbf{b} show the modified $\beta$ profile with a constant offset in the deep and shallow cases, respectively. \textbf{c}-\textbf{f} are the same as Fig.~\ref{fig:Mechanism_compare} but for simulations with the modified $\beta$ profiles. The simulation parameters are $L_d \to \infty$, $\varepsilon=5\times10^{-7}$, and $\alpha=5\times10^{-5}$.}\label{fig:S3T}
\end{figure*}

\begin{figure*}[b]
\centering
\includegraphics[width=1.0\textwidth]{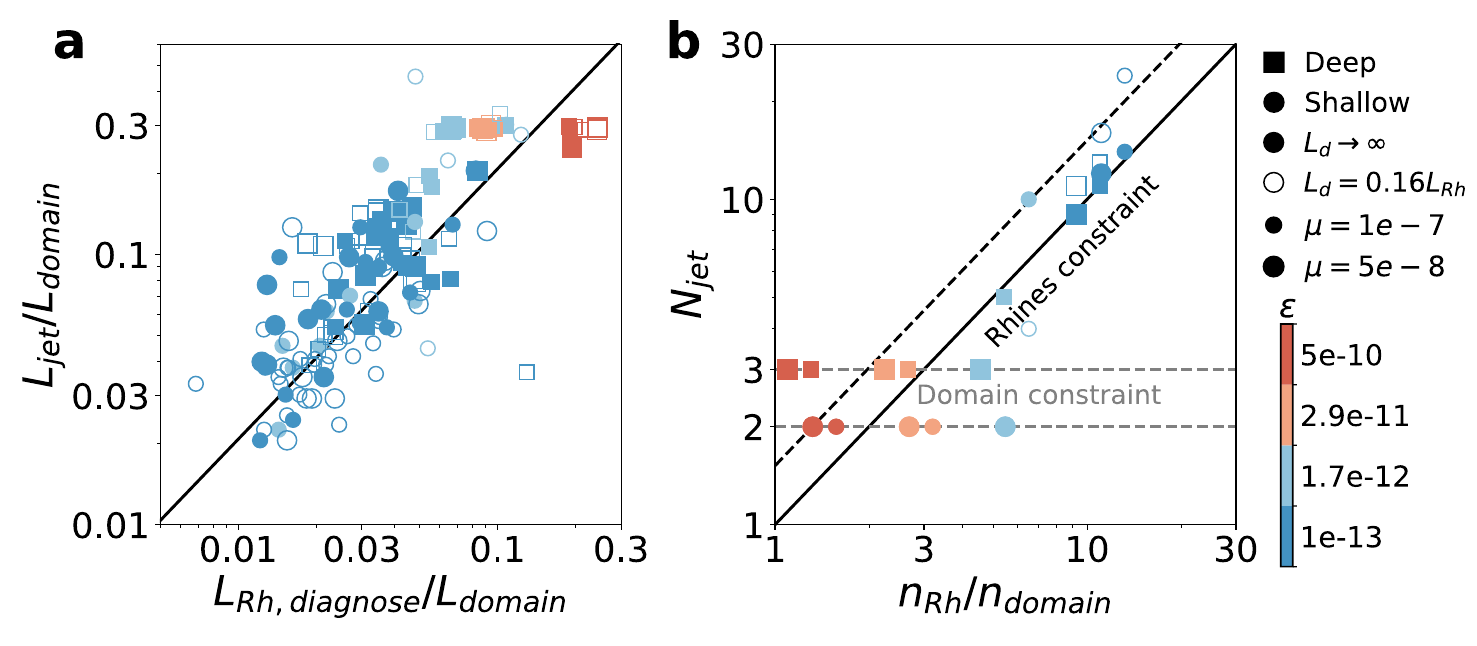}
\caption{\textbf{Jet width and number.} \textbf{a}, Diagnosed jet width ($L_{jet}$) compared to the diagnosed modified Rhines scale ($L_{Rh,diagnose}$), both normalized by the domain length scale $L_{domain}$. \textbf{b}, Number of jets ($N_{jet}$) diagnosed in the simulations compared to the theoretical prediction, $n_{Rh}/n_{domain}$. Detailed jet diagnostic information is provided in APPENDIX~\ref{app:results}. In \textbf{a}, jets with widths of $1/2$ the domain size in the shallow scenario and $1/3$ the domain size in the deep scenario are excluded, as their widths are constrained by the domain size and are not expected to follow Rhines scaling. The black solid line represents a fitted line. In \textbf{b}, the black solid line is the one-to-one line, and the black dashed line indicates the adjusted prediction, incorporating the fitted pre-factor (2.07) of jet width to Rhines scale from \textbf{a}. Gray dashed lines represent predictions where the jet scale is constrained by the domain size: $N_{jet} = 3$ for the deep scenario and $N_{jet} = 2$ for the shallow scenario. In \textbf{a} and \textbf{b}, regions outside the TC in the deep scenario are excluded.}\label{fig:Rhines}
\end{figure*}

The results from the numerical simulations show that the direction of the equatorial jet is completely controlled by the $\beta$ profile, which is determined by the spherical geometry. In all deep-scenario simulations, a prograde jet forms at the equator; conversely, in all shallow-scenario simulations, a retrograde jet forms at the equator (Fig.~\ref{fig:direction}). This behavior is consistent with both PV mixing and SSST. With $\beta$ being negative near the equator in the deep case, a prograde jet should form if PV is to be homogenized. The opposite is true for the shallow case. Evidence of relative vorticity (blue) compensating planetary vorticity (red) can be seen from Fig.~\ref{fig:Mechanism_compare}c and d, but the compensation is far from perfect. 

SSST theory can also correctly predict the jet direction given the $\beta$ gradient near the equator. Using the prescribed $\beta$ profile, we compute the eddy momentum flux convergence $\overline{v q}=-\partial \overline{uv}/\partial y$ predicted by SSST (Fig.~\ref{fig:Mechanism_compare}a-b; see APPENDIX~\ref{app:S3T} for details). The sign of the equatorial acceleration agrees with the simulations, indicating that the eddy-driven acceleration can determine the direction of the equatorial jet. The resulting jet profile does not exactly match the diagnosed acceleration because this calculation isolates the effect of the $\beta$ profile alone, without accounting for feedback from the zonal jets or nonlinear eddy-eddy interactions, which may further adjust the jet profile. To test whether multiple equilibria exist that favor both prograde and retrograde equatorial jets, we repeated two deep simulations initialized with a retrograde equatorial jet and two shallow simulations initialized with a prograde equatorial jet. All experiments end up reversing their equatorial jets (Fig.~\ref{figS2:IC}). In addition, in simulations with strong energy injection and strong damping (e.g., those shown in Fig.~\ref{figS2:IC}), poleward jet migration is observed \citep{chemke2015poleward}. In contrast, in simulations with weaker energy injection and damping, the jets remain relatively steady and no systematic migration occurs.

In these simulations, both PV homogenization and SSST appear able to predict the jet direction. This is because in realistic giant planet atmospheres, the sign and gradient of $\beta$ typically change simultaneously when transitioning between deep and shallow profiles (Fig.~\ref{fig:sketch-beta-h}b~\&~d), making their effects difficult to distinguish. To separate these effects, we construct modified $\beta$ profiles by adding a constant offset to the deep and shallow cases, such that $\beta$ becomes entirely positive or negative while preserving its gradient (Fig.~\ref{fig:S3T}a-b). The shifted profiles still produce a prograde equatorial jet in the deep case and a retrograde jet in the shallow case, consistent with the predicted eddy momentum flux convergence by SSST (Fig.~\ref{fig:S3T}c-d). However, the gradient of relative vorticity now takes the same sign as the gradient of planetary vorticity near the equator (Fig.~\ref{fig:S3T}e-f), so the two components enhance instead of compensating one another, contradicting the PV mixing mechanism. This result indicates that in our simulations, statistically averaged eddy fluxes controlled by the $\beta$ gradient play the dominant role in setting the equatorial jet direction rather than the absolute sign of $\beta$.

Our results further indicate that the direction of the equatorial jet is closely related to the geometry of the system through the jet penetration depth $D$, which shapes the $\beta$ profile. As $D$ increases, the equatorial jet transitions from retrograde to prograde. Importantly, it should be noted that all giant planet atmospheres contain regions outside the TC. The distinction between deep and shallow regimes should therefore be understood in terms of the relative scale between the equatorial jet width, characterized by the equatorial Rhines scale $L_{Rh,eq}$, and the jet penetration depth $D$. When $L_{Rh,eq}/D \lesssim 1$, the region outside the TC remains dynamically significant for the equatorial jet, corresponding to a deep scenario. In contrast, when $L_{Rh,eq}/D \gg 1$, the equatorial jet becomes sufficiently broad relative to the penetration depth, and the system enters a shallow scenario. In this limit, modes associated with the negative $\beta$ outside the TC primarily redistribute momentum within the jets rather than control momentum exchange between the equatorial jet and higher latitudes, and therefore play a weaker role in setting the equatorial jet direction.

Similar geometric influences have also been proposed in previous studies.  \cite{camisassa2022solar} highlighted the importance of the ratio between the jet penetration depth and the width of the convection cells. They showed that when the atmosphere is shallower than the convection cell width, the convective Rossby number rises above unity, disrupting the tilted Busse mode and reversing the equatorial jet from prograde to retrograde. \cite{duer2025gas} suggested that the equatorial Busse mode is directly influenced by the jet penetration depth. As the penetration depth decreases, the Busse mode may reverse its tilt, causing the equatorial jet to switch from prograde to retrograde and leading to bistability. Although the transient structures discussed here differ from the Busse modes considered in these studies, both mechanisms share a common geometric origin: the structure and tilt of the dynamically relevant modes, and therefore the direction of the equatorial jet, are strongly influenced by the gradient of the $\beta$ profile imposed by spherical geometry.

\section{The number of jets}\label{sec:number}

The TC divides the planetary atmosphere into two distinct regions. On Jupiter and Saturn, while multiple jets can form inside the TC, the region outside the TC only has one single jet. We believe this occurs because the atmosphere outside the TC is not in direct contact with the Ohmic dissipation layer, and therefore cannot be damped as effectively as elsewhere. Without damping, jet energy accumulates, leading to an increase in jet velocity, along with an expansion of jet width following the Rhines scale \citep{rhines1975waves}. Once the jet spans the entire region outside the TC, it comes into contact with the Ohmic dissipation layer, which subsequently limits further growth.

The jet width inside the TC follows the Rhines scale, $L_{Rh}$ \citep{rhines1975waves}. In an unstratified atmosphere, the Rhines scale is given by $L_{Rh} \sim (U/\beta)^{1/2}$. When stratification is present, it modifies the Rhines scale to $L_{Rh} \sim (\beta/U - L_d^{-2})^{-1/2}$, where $L_d$ is the deformation radius \citep{theiss2006generalized,slavin2012multiple,di2013simulating}. The modified Rhines scale captures the jet width in our numerical simulations fairly well (Fig.~\ref{fig:Rhines}a). The number of jets, in turn, scales with the ratio of the domain scale to the Rhines scale (Fig.~\ref{fig:Rhines}b), consistent with findings from previous studies \citep[e.g.,][]{read2004jupiter,sukoriansky2007arrest,dunkerton2008barotropic,read2015experimental,lonner2022planetary,lemasquerier2023zonal}. In high-energy experiments (denoted by the reddish color in Fig.~\ref{fig:Rhines}b), the jet number scaling breaks down as the jet width becomes sufficiently large to feel the constraints of the domain. In shallow scenario simulations, one prograde jet must form together with the equatorial retrograde jet by virtue of momentum conservation, resulting in a total of two jets. In the deep scenario, the TC further divides the domain into two regions, leading to the formation of three jets, excluding the equatorial one (Fig.~\ref{fig:Rhines}b).

\section{The smoothness of jets}\label{sec:smoothness}

\begin{figure*}[b]
\centering
\includegraphics[width=0.9\textwidth]{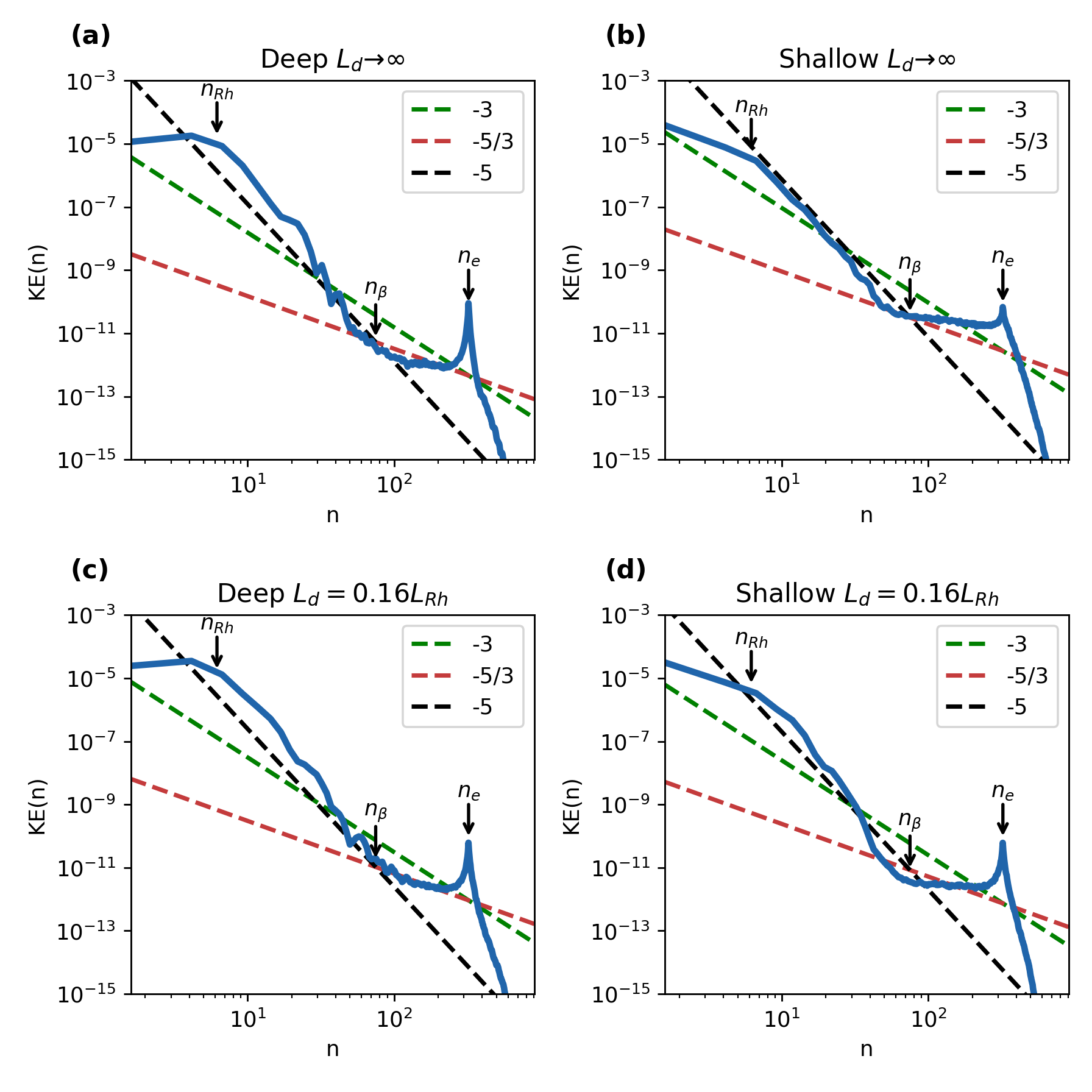}
\caption{\textbf{Kinetic energy spectrum.} Snapshots of the kinetic energy (KE; $u^2/2 + v^2/2$) spectra from simulations with $\varepsilon=2.9 \times10^{-11}$ and $\mu=10^{-7}$. The blue line shows the simulated KE spectrum. The green, red, and black dashed lines indicate the theoretical slopes of $-3$, $-5/3$, and $-5$, respectively. The transition wavenumbers $n_e$, $n_\beta$, and $n_{Rh}$ are also marked, where $n_\beta$ and $n_{Rh}$ are estimated using the averaged $\beta$ across the domain.}
\label{figS:spectrum}
\end{figure*}

\begin{figure*}[t]
\centering
\includegraphics[width=0.6\textwidth]{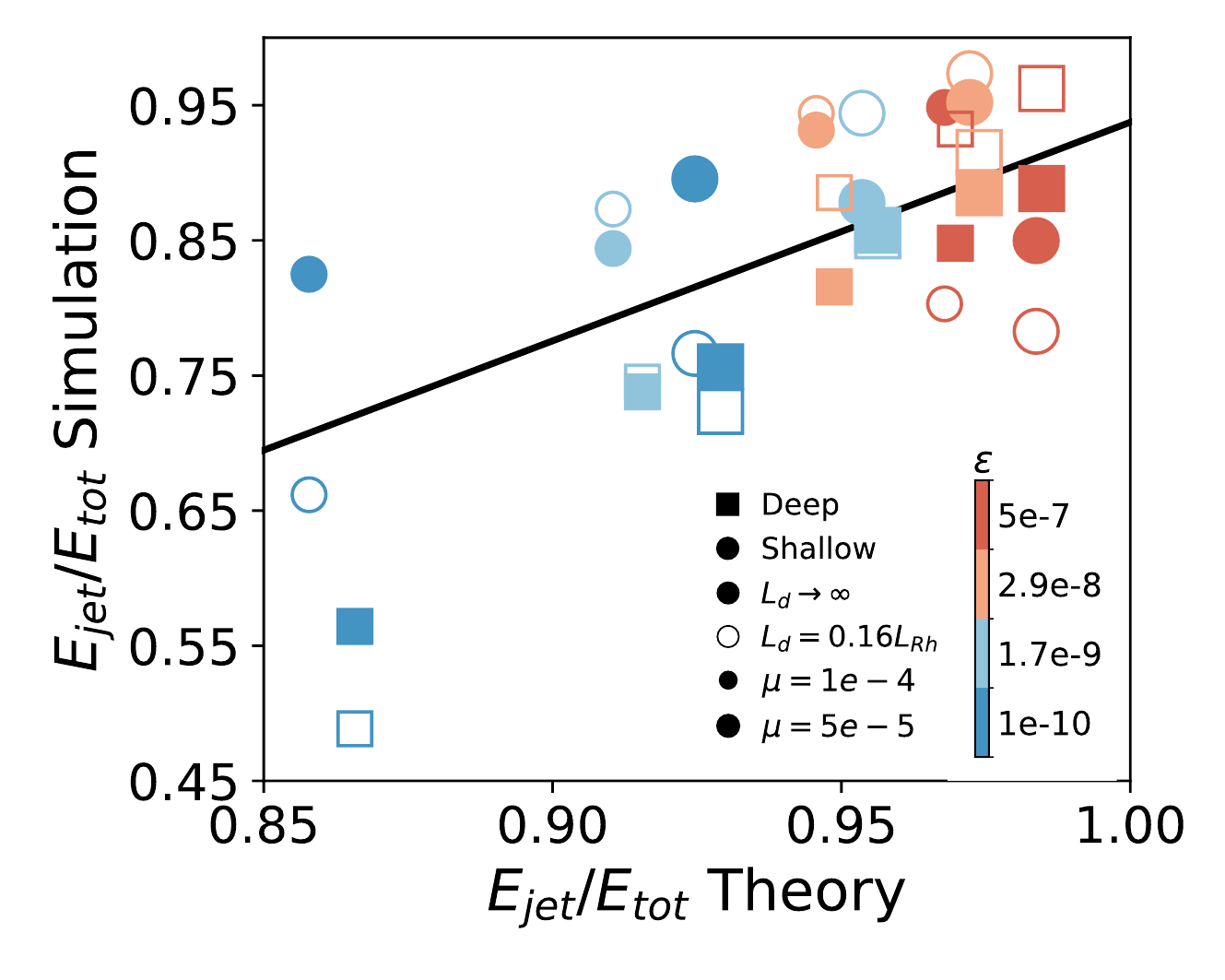}
\caption{\textbf{The smoothness of the jets.} The jet energy fraction ($E_{jet}/E_{tot}$) diagnosed in the simulations is compared with predictions from Equation~\ref{eq:Ejet_to_Etot}. Results are shown for the equilibrium state simulations with higher energy injection rates and damping rates ($\varepsilon=5 \times 10^{-7}$, $2.9 \times 10^{-8}$, $1.7 \times 10^{-9}$, and $10^{-10}$; $\mu=5 \times 10^{-5}$ and $10^{-4}$), as simulations with lower injection and damping rates have not reached full equilibrium. The black line represents a linear fit.}
\label{figS4:Energy}
\end{figure*}

In a 2-D fluid system, small-scale eddies tend to organize themselves into larger coherent structures through a process known as the inverse energy cascade\footnote{It should be noted that although SSST describes a non-local transfer of energy from small-scale eddies directly to large-scale jets, our system is fully nonlinear, allowing eddy-eddy interactions and associated energy cascades to occur simultaneously.}. After the energy cascades to a characteristic scale $L_\beta$, where the influence of the planetary PV gradient ($\beta$) becomes significant, the eddies transfer their energy into zonal jets. Beyond that, energy continues cascading to larger scales until it reaches the frictional Rhines scale $L_{Rh}$, where it is dissipated \citep{maltrud1991energy,vallis1993generation}. If there is a sufficiently large scale separation among the scale of the injected eddies $L_e$, the $\beta$ scale $L_\beta$, the frictional Rhines scale $L_{Rh}$, and the domain scale $L_{domain}$, the flow enters the zonostrophic regime where multiple zonal jets can form \citep[e.g.,][]{galperin2008zonostrophic,cabanes2017laboratory,lemasquerier2023zonal}. The criterion is given by $n_e \gtrsim 4 n_\beta \gtrsim 8 n_{Rh} \gtrsim 30 n_{domain}$, where $n$ indicates the corresponding wavenumbers, with $n_\beta = \varepsilon^{-1/5} \beta^{3/5}/2$ and $n_{Rh} = \varepsilon^{-1/4} \beta^{1/2} \mu^{1/4}/\sqrt{2}$, following \cite{galperin2008zonostrophic}. $n_e = \pi / L_e$ and $n_{domain} = \pi / L_{domain}$ are the eddy and domain wavenumbers, respectively.

The giant planet atmosphere is likely in this zonostrophic regime with clear scale separation (Table~\ref{tab:planet_parameter}). In this regime, the kinetic energy of zonal jets peaks at larger scales, while the kinetic energy of eddies is primarily concentrated at smaller scales. When more energy is stored in the jets, they appear smoother; conversely, when eddies hold more energy, the jets look more chaotic \citep{scott2012structure}. We can determine the energy distribution between jets and eddies by integrating the theoretical energy spectrum from $n_{Rh}$ to $n_\beta$ and from $n_\beta$ to infinity (infinitely small length scale). For a 2-D rotating system forced by small-scale eddies, the energy spectrum exhibits different slopes across various scales \citep{vallis1993generation}: an $n^{-3}$ slope for scales smaller than the forcing scale, where enstrophy cascades to smaller scales and is dissipated by viscosity; an $n^{-5/3}$ slope in the inertial range between the forcing scale and the $\beta$ scale, where energy inversely cascades to larger scales; and an $n^{-5}$ slope for scales between the $\beta$ scale and the frictional Rhines scale, where eddies cascade their energy into zonal jets, eventually dissipated by large-scale damping. The spectrum is given by

\begin{equation}\label{eq:kspectrum}
E(n) =
\begin{cases} 
C_E n_\beta^{10/3} n^{-5},  & \mathrm{if}\ n_{Rh} < n < n_\beta, \\
C_E n^{-5/3},  & \mathrm{if}\ n_\beta < n < n_e, \\
C_E n_e^{4/3} n^{-3},  & \mathrm{if}\ n_e< n < n_\nu,
\end{cases}
\end{equation}

\noindent where $C_E$ is a constant determined by the total energy and $n_\nu$ is the scale where small-scale viscous dissipation dominates. Assuming that the viscous scale is very small and negligible ($n_\nu \to \infty$), the ratio of the jet energy to the total energy is given by

\begin{equation}\label{eq:Ejet_to_Etot}
\begin{aligned}
\frac{E_{jet}}{E_{tot}} &= \frac{ \int_{n_{Rh}}^{n_\beta} C_E n_\beta^{10/3} n^{-5} dn }{ \int_{n_{Rh}}^{n_\beta} C_E n_\beta^{10/3} n^{-5} dn + \int_{n_\beta}^{n_e} C_E n^{-5/3} dn + \int_{n_e}^{\infty} C_E n_e^{4/3} n^{-3} dn}  \\
&= \frac{n_\beta^{10/3}(n_{Rh}^{-4}-n_\beta^{-4})}{n_\beta^{10/3}(n_{Rh}^{-4}-n_\beta^{-4}) + 6(n_\beta^{-2/3}-n_e^{-2/3}) + 2n_e^{-2/3}} \\
&= \frac{(n_\beta/n_{Rh})^{4}-1}{(n_\beta/n_{Rh})^{4}+5-4(n_\beta/n_{e})^{2/3}}.
\end{aligned}
\end{equation}

In our simulations, the inverse energy cascade is well represented, where both the $-5/3$ and $-5$ slopes are approximately captured, with a spectral peak at the forcing wavenumber $n_e$ (Fig.~\ref{figS:spectrum}). This feature is less well captured in some deep cases, likely because the domain includes regions both inside and outside the TC, which exhibit different jet characteristics and contain a sharp $\beta$ jump at the TC (APPENDIX~\ref{app:results}). At scales smaller than the forcing scale, however, the spectrum becomes much steeper than the expected $-3$ slope, likely due to strong hyperviscous dissipation at small scales introduced to maintain numerical stability. We further examine the energy fraction using simulations with higher energy injection and damping rates, since the cases with smaller rates have not yet reached a fully equilibrated state. The estimate qualitatively predicts the outcomes observed in numerical simulations (Fig.~\ref{figS4:Energy}).

The theoretical estimate (Equation~\ref{eq:Ejet_to_Etot}) shows that $E_{jet}/E_{tot}$ is primarily influenced by the zonostrophic parameter, $n_{\beta}/n_{Rh}$ \citep{galperin2006anisotropic,sukoriansky2007arrest}, assuming that $n_e$ is significantly larger than $n_{\beta}$ (Table~\ref{tab:planet_parameter}). A greater separation between $n_\beta$ and $n_{Rh}$ leads to more energy being stored in jets rather than eddies, making this ratio a useful qualitative metric for predicting the jet energy fraction. By substituting the definitions of $n_\beta$ and $n_{Rh}$, we find that $n_\beta/n_{Rh} \sim \varepsilon^{1/20}\beta^{1/10} \mu^{-1/4}$. Consequently, a higher energy injection rate $\varepsilon$, a higher $\beta$ value (indicating either faster rotation rate or smaller planetary radius), or a weaker large-scale damping $\mu$, result in more energy concentrated in the jets, making them smoother. Within our simulation parameters, the deformation radius does not significantly affect the jet energy fraction.

\section{Conclusion and discussion}\label{sec:discussion}

Here, we use a stochastically forced 2-D Quasi-Geostrophic framework to study the factors controlling the direction, number, and smoothness of zonal jets in the atmospheres of giant planets. Our results show that the direction of the equatorial jet primarily depends on the $\beta$ profile, which in turn is determined by the jet penetration depth. When the penetration depth is large, a prograde jet forms at the equator, whereas when the penetration depth is small\footnote{Note that the $\beta$ profile in our shallow cases resembles that in previous studies on Rossby wave propagation mechanisms \citep[e.g.,][]{schneider2009formation,liu2010mechanisms}, which consider a shallow, stratified atmosphere.}, the equatorial jet becomes retrograde. We further find that the equatorial jet direction in our simulations is controlled mainly by the gradient of $\beta$, which regulates the transient growth and decay of eddies as described by SSST \cite[e.g.,][]{farrell2003structural}, rather than the absolute sign of $\beta$, as suggested by the PV mixing mechanism \citep[e.g.,][]{yano2005deep,vallis2019essentials}.

This result provides insight into the interior structure of the giant planets. On Jupiter and Saturn, the equatorial jet widths are comparable to the penetration depths ($L_{Rh,eq}/D\sim 1$, Table~\ref{tab:planet_parameter}), indicating a dynamically deep regime that favors prograde equatorial jets, consistent with observations. This equatorial jet outside the TC is also stronger than the jets formed at higher latitudes due to the lack of damping (Fig.~\ref{fig:direction}b), also consistent with observations (Fig.~\ref{fig:u-observation}a). For Uranus and Neptune, although the estimates of the jet penetration depth remain uncertain, their observed retrograde equatorial jets appear incompatible with a deep geometry and low convective Rossby number, instead aligning more closely with a shallow geometry scenario ($L_{Rh,eq}/D \gg 1$, Table~\ref{tab:planet_parameter}), suggesting the presence of either an Ohmic dissipation layer or a stably stratified layer near the surface \citep{wulff2022zonal,wulff2024effects,christensen2024quenching}. In a shallow system, a retrograde equatorial jet forms, and the absence of a region outside the TC would prevent the shallow system from forming an excessively strong equatorial jet relative to the high-latitude jets (Fig.~\ref{fig:direction}c). Both features are consistent with observations (Fig.~\ref{fig:u-observation}b). 

In our simulations, the simulated jet width follows the Rhines scale, consistent with previous numerical and laboratory studies \citep[e.g.,][]{read2004jupiter,sukoriansky2007arrest,dunkerton2008barotropic,read2015experimental,lonner2022planetary,lemasquerier2023zonal}. This scaling suggests that the number of jets is well predicted by the ratio of the domain size (planetary radius for a hemisphere) to the Rhines scale, which can be estimated from the observable characteristic flow speed. When the Rhines scale is comparable to the domain size, only one single wide jet forms at mid- and high-latitudes. On Jupiter and Saturn, multiple alternating jets are observed at mid-latitudes, consistent with our theoretical predictions. On Uranus and Neptune, there is only one prominent prograde jet in each hemisphere outside the equatorial region, which also aligns with our theoretical prediction (Fig.~\ref{fig:direction}~\&~Fig.~\ref{fig:Rhines}b). The mechanism that sustains such strong jets on Uranus and Neptune remains unclear and requires further investigation.

Lastly, our results show that the energy partition between jets and eddies is governed by the zonostrophic parameter $n_\beta/n_{Rh}$ \citep{galperin2006anisotropic,sukoriansky2007arrest}. When $n_\beta/n_{Rh}$ is large, i.e., on planets with higher energy input, faster rotation, smaller radius, or weaker large-scale damping, a greater fraction of energy resides in zonal jets rather than eddies, resulting in smoother jet structures.

The interpretation that the equatorial jet direction is controlled by the gradient of $\beta$, rather than its absolute sign, reflects the behavior observed in our simulations and may not be generally applicable across different parameter regimes. Therefore, it should not be taken as the only possible interpretation of jet direction on giant planets. In particular, the PV mixing mechanism, in which the jet direction is determined by the sign of $\beta$, is also consistent with our simulations using realistic $\beta$ profiles (Fig.~\ref{fig:direction}c-d). Determining the parameter regimes in which different mechanisms dominate, including the Busse mode and Rossby wave propagation, remains an open question.

We consider an atmosphere that is barotropic along the rotation axis and introduce a deformation radius $L_d$ into the model to represent the coupling between the dynamical atmosphere and the Ohmic dissipation layer (see APPENDIX~\ref{app:model}). Near the surface, solar heating generates a stably stratified weather layer, where the atmosphere is baroclinic, as evidenced by observations from the Galileo Probe Doppler Wind Experiment \citep{atkinson1998galileo}. To account for these effects, we inject the vorticity anomalies induced by baroclinic instability directly as stochastic noise. Additionally, waves originating in deeper layers may propagate upward, dissipate, and deposit momentum, thereby contributing to long-term variability in jet strength \citep{leovy1991quasiquadrennial,fouchet2008equatorial,showman2019atmospheric}. A more comprehensive investigation of how baroclinicity modifies jet structure is left for future work. Moreover, our use of a Cartesian coordinate neglects curvature effects. Extending the analysis to a curvilinear coordinate would help assess the robustness of the mechanisms identified here under more realistic geometries.

Despite these potential complexities, our results demonstrate that jet penetration depth can strongly influence surface wind patterns, suggesting that the interior structures of giant planets may be inferred from surface observations alone. These projections can be validated through improved characterization of the gravity field by future missions. Our 2-D framework allows direct comparison of deep and shallow scenarios within a unified setting and enables exploration of a broad parameter space with modest computational cost. This framework may serve as a foundation for future studies that incorporate additional physical processes to further our understanding of jet dynamics on giant planets.

\begin{acknowledgments}
We thank Keaton Burns and Daniel Lecoanet for their assistance in setting up the Dedalus numerical model. We thank Yohai Kaspi, Philip Marcus, Jiaru Shi, Huazhi Ge, and two anonymous reviewers for helpful discussions and comments. This work originates from the 2023 Geophysical Fluid Dynamics (GFD) Summer School at Woods Hole. We thank the GFD faculty for organizing the program and the Woods Hole Oceanographic Institution for hosting it. The numerical simulations were performed using the svante cluster in the Department of Earth, Atmospheric, and Planetary Sciences (EAPS) at MIT. We thank Jeffery Scott and Morgan Ludwig for their technical support. For the purpose of open access, the authors have applied a Creative Commons Attribution (CC BY) licence to any Author Accepted Manuscript version arising from this submission.
\end{acknowledgments}

\begin{contribution}

Y.Z., W.K., G.F., and G.V. designed the research. Y.Z. performed research and analyzed data. Y.Z. and W.K. wrote the first manuscript, and all authors contributed to revising the manuscript.

\end{contribution}


\newpage

\appendix

\section{Detailed model description}\label{app:model}

Following the derivation in \citet{yano1994jupiter}, we asymptotically expand the momentum and mass continuity equations of an isentropic anelastic fluid in powers of the Rossby number, $\Ro$, and apply the quasi-geostrophic approximation. The momentum equation for an anelastic fluid is \citep{kaspi2009deep}:

\begin{equation}\label{Eq:Anelastic}
    \frac{D \boldsymbol{v}}{\partial t} + 2\boldsymbol{\Omega} \times \boldsymbol{v} = -\nabla P' + \alpha_s s' \nabla \Phi,
\end{equation}

\noindent where $\boldsymbol{v}=u\boldsymbol{e_x}+v\boldsymbol{e_y}+w\boldsymbol{e_z}$ is the 3-D velocity (coordinate defined in Fig.~\ref{fig:sketch-beta-h}), $\boldsymbol{\Omega}$ is the planetary rotation, $P'=p'/\tilde{\rho}$, $\nabla \Phi = - \nabla \tilde{p}/\tilde{\rho}$, $p$ is the pressure, and $\rho$ is the density, with tilde for background state and prime for perturbation. $s'$ is the entropy anomaly, and $\alpha_s \equiv -(1/\tilde{\rho})(\partial \rho / \partial s)_p$, where the subscript $p$ indicates differentiation at constant pressure. We assume that the atmosphere is isentropic ($s'=0$), corresponding to strong turbulent mixing in the deep convective layer, and neglect entropy anomalies associated with surface stratification and solar forcing. Let the typical velocity and length scales be $U$ and $L$, respectively, and assume $P' \sim 2\Omega U L$. The non-dimensional momentum equations are

\begin{equation}
    \Ro \frac{D \hat{u}}{D \hat{t}} - \hat{v} = -\frac{\partial \hat{P}}{\partial \hat{x}},
\end{equation}

\begin{equation}
    \Ro \frac{D \hat{v}}{D \hat{t}} + \hat{u} = -\frac{\partial \hat{P}}{\partial \hat{y}},
\end{equation}

\begin{equation}
    \Ro \frac{D \hat{w}}{D \hat{t}} = -\frac{\partial \hat{P}}{\partial \hat{z}},
\end{equation}

\noindent where hat indicates non-dimensional variables and $\Ro=U/(2\Omega L)$ is the Rossby number. The non-dimensional mass continuity equation is

\begin{equation}
    \hat{\nabla} \cdot \boldsymbol{\hat{v}} + \lambda \hat{v}_r = \hat{\nabla} \cdot \boldsymbol{\hat{v}} + \lambda (-\hat{v}\cos{\phi}+\hat{w}\sin{\phi}) = 0,
\end{equation}

\noindent where $\lambda = (1/\tilde{\rho})(d\tilde{\rho}/d\hat{r})$ represents stratification, $r$ is radius, $v_r$ is the radial velocity, and $\phi$ is latitude. For rapidly rotating planets, we assume $\Ro \ll 1$ and expand each variable in powers of $\Ro$, e.g., $\hat{u}=\hat{u}_0+\Ro \hat{u}_1+\dots$. The leading-order momentum balance then gives

\begin{equation}\label{eq:geo+hydro}
    \hat{u}_0 = -\frac{\partial \hat{P}_0}{\partial y}, \ \hat{v}_0 = \frac{\partial \hat{P}_0}{\partial x}, \ \frac{\partial \hat{P}_0}{\partial z}=0,
\end{equation}

\noindent which gives hydrostatic and geostrophic balance. Since the leading-order horizontal velocity is divergent-free, we may thus introduce a streamfunction $\hat{\Psi}_0=\hat{P}_0$ to represent it as $\hat{u}_0=-\partial \hat{\Psi}_0/\partial \hat{y}$, $\hat{v}_0=\partial \hat{\Psi}_0/\partial \hat{x}$. Assuming $\hat{w}_0=0$, the zeroth-order mass continuity equation becomes

\begin{equation}
    \lambda \cos{\phi} \hat{v}_0=0.
\end{equation}

\noindent Since $\hat{v}_0$ is not zero everywhere, this term must appear at the next order equation, implying $\lambda \cos{\phi} \sim \mathrm{O}(\Ro)$. This requires either weak stratification or proximity to the pole (Fig.~\ref{fig:sketch-beta-h}). The first-order equations are

\begin{equation}\label{eqRo1:u}
    \frac{D_0 \hat{u}_0}{D \hat{t}} - \hat{v}_1 = -\frac{\partial \hat{P}_1}{\partial \hat{x}},
\end{equation}

\begin{equation}\label{eqRo1:v}
    \frac{D_0 \hat{v}_0}{D t} + \hat{u}_1 = -\frac{\partial \hat{P}_1}{\partial \hat{y}},
\end{equation}

\begin{equation}\label{eqRo1:w}
    \frac{D_0 \hat{w}_0}{D t} = -\frac{\partial \hat{P}_1}{\partial \hat{z}},
\end{equation}

\begin{equation}\label{eqRo1:mass}
    \frac{\partial \hat{u}_1}{\partial \hat{x}} + \frac{\partial \hat{v}_1}{\partial \hat{y}} + \frac{1}{\tilde{\rho}} \frac{\partial (\tilde{\rho} \hat{w}_1)}{\partial \hat{z}} - \frac{\lambda \cos{\phi}}{\mathrm{Ro}} \hat{v}_0 = 0,
\end{equation}

\noindent where $D_0/D\hat{t} \equiv \partial/\partial \hat{t} + \hat{u}_0 \partial/\partial \hat{x} + \hat{v}_0 \partial/\partial \hat{y}$. Taking $\partial (\ref{eqRo1:v})/\partial \hat{x} - \partial (\ref{eqRo1:u})/\partial \hat{y}$, we have

\begin{equation}
    \frac{D_0}{D \hat{t}} \left( \frac{\partial \hat{v}_0}{\partial \hat{x}} - \frac{\partial \hat{u}_0}{\partial \hat{y}} \right) + \left( \frac{\partial \hat{u}_0}{\partial \hat{x}} + \frac{\partial \hat{v}_0}{\partial \hat{y}} \right) \left( \frac{\partial \hat{v}_0}{\partial \hat{x}} - \frac{\partial \hat{u}_0}{\partial \hat{y}} \right) + \frac{\partial \hat{u}_1}{\partial \hat{x}} + \frac{\partial \hat{v}_1}{\partial \hat{y}} = - \frac{\partial}{\partial \hat{y}} \left( \frac{\partial \hat{P}_1}{\partial \hat{x}} \right) + \frac{\partial}{\partial \hat{x}} \left( \frac{\partial \hat{P}_1}{\partial \hat{y}} \right) = 0.
\end{equation}

\noindent Note that $\partial \hat{u}_0/\partial \hat{x}+\partial \hat{v}_0/\partial \hat{y}=0$, and substitute Equation~\ref{eqRo1:mass} to eliminate $\partial \hat{u}_1/\partial \hat{x}+\partial \hat{v}_1/\partial \hat{y}$, we obtain the vorticity equation:

\begin{equation}\label{eqRo1:vort}
    \frac{D_0}{D \hat{t}} \hat{\nabla}^2 \hat{\Psi}_0 = \frac{1}{\tilde{\rho}} \frac{\partial (\tilde{\rho} \hat{w}_1)}{\partial \hat{z}} - \frac{\lambda \cos{\phi}}{\mathrm{Ro}} \hat{v}_0,
\end{equation}

\noindent where we used the fact that $\partial \hat{v}_0/\partial \hat{x}-\partial \hat{u}_0/\partial \hat{y}=\hat{\nabla}^2 \hat{\Psi}_0$.

On rapidly rotating giant planets, motions are largely invariant along the rotation axis. Assuming the flow is barotropic and integrating the density-weighted vorticity equation along this axis, $\int_{z_b}^{z_t} \tilde{\rho} (\ref{eqRo1:vort}) dz$, we obtain

\begin{equation}
    \hat{H}_\rho \frac{D_0 }{D \hat{t}} \hat{\nabla}^2 \hat{\Psi}_0 = (\tilde{\rho} \hat{w}_1)|^{\hat{z}_t}_{\hat{z}_b} - \frac{\hat{v}_0}{\Ro} \int_{\hat{z}_b}^{\hat{z}_t} \cos{\phi} \frac{d\tilde{\rho}}{d \hat{r}} d\hat{z}.
\end{equation}

Note that 

\begin{equation}
    \int_{\hat{z}_b}^{\hat{z}_t} -\cos{\phi}\frac{d\tilde{\rho}}{d \hat{r}} d\hat{z} = \int_{\hat{z}_b}^{\hat{z}_t}  \frac{\partial \tilde{\rho}}{\partial \hat{y}} d\hat{z} = \frac{d}{d\hat{y}} \left(\int_{\hat{z}_b}^{\hat{z}_t} \tilde{\rho} d\hat{z} \right) - \tilde{\rho}(\hat{z}_t) \frac{d \hat{z}_t}{d\hat{y}} + \tilde{\rho}(\hat{z}_b) \frac{d \hat{z}_b}{d\hat{y}} = \frac{d}{d\hat{y}} \left(\int_{\hat{z}_b}^{\hat{z}_t} \tilde{\rho} d\hat{z} \right) - (\tilde{\rho} \cot{\phi})|^{\hat{z}_t}_{\hat{z}_b}.
\end{equation}

\noindent Restoring the dimensional form and noting that $w-v\cot{\phi}=v_r/\sin{\phi}$, we obtain

\begin{equation}\label{eq:yanoPV}
    \frac{D \nabla^2 \psi}{Dt} + \beta \frac{\partial \psi}{\partial x} = \frac{2\Omega}{H_\rho} \left(\tilde{\rho} \frac{v_r}{\sin{\phi}} \right) \bigg|_{z_b}^{z_t},
\end{equation}

\noindent where $\psi$ is the streamfunction, satisfying $u = -\partial \psi/\partial y$ and $v = \partial \psi/\partial x$, $D/Dt \equiv \partial/\partial t + u \partial/\partial x + v \partial/\partial y$, and $H_\rho(y) = \int_{z_b(y)}^{z_t(y)} \tilde{\rho} dz$ is the density-weighted column depth. Here $z_b$ and $z_t$ denote the lower and upper boundaries of the atmospheric column, respectively. The $\beta$ term is given by $\beta = -(2\Omega/H_\rho)(dH_\rho/dy)$, arising from the effects of density stratification and topography, consistent with prior formulations by \cite{ingersoll1982motion}, \cite{yano1994jupiter}, and \cite{gastine2014zonal}\footnote{Note that in our model, the $y$ direction points inward in the northern hemisphere, opposite to the outward radial direction in conventional cylindrical coordinates.}. 

\begin{figure*}[t]
\centering
\includegraphics[width=0.5\textwidth]{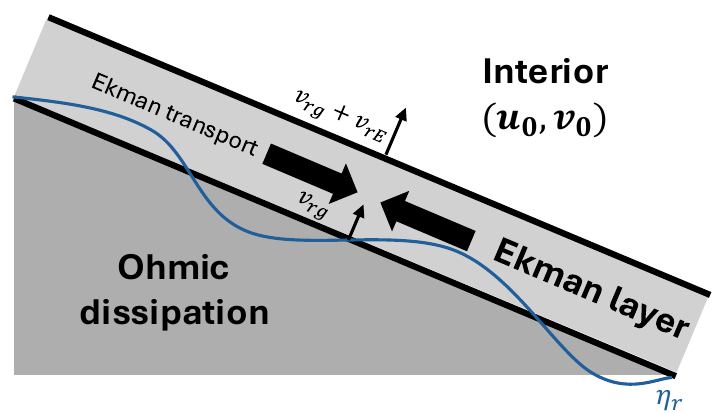}
\caption{\textbf{Sketch of the Ekman boundary layer.} The radial velocity at the bottom boundary ($v_r(z_b)$) has two components: $v_{rg}$, associated with the interface perturbation ($\eta_r$), and $v_{rE}$, associated with the convergence of Ekman transport (thick black arrows). See text for details.}
\label{figS:Ekman}
\end{figure*}

At the top of the atmosphere, the density vanishes ($\tilde{\rho} \approx 0$). At the bottom boundary, the radial velocity $v_r(z_b)$ has two components: one arising from perturbations of the interface between the dynamical interior and the underlying Ohmic layer, $v_{rg}$, and another from the convergence of transport within the viscous boundary layer, $v_{rE}$ (Fig.~\ref{figS:Ekman}). Ohmic dissipation damps the flow to nearly zero, which can be represented by a no-slip bottom boundary. Due to viscosity, an Ekman boundary layer forms on a rotating planet, connecting the no-slip lower boundary with the overlying moving interior. Horizontal convergence associated with the Ekman boundary flows (the Ekman transport) induces an vertical velocity at the boundary layer top, $v_{rE}$, expressed as (see Equation~4.9.36 in \cite{pedlosky2013geophysical})

\begin{equation}
    v_{rE} = w \sin{\phi} - v\cos{\phi} = \sqrt{\frac{\nu_V}{4\Omega}} \nabla^2 \psi \sin{\phi},
\end{equation}

\noindent assuming the bottom slope is small, where $\nu_V$ is the vertical viscosity and $\sqrt{\nu_V/(4\Omega)}$ characterizes the Ekman boundary layer depth. The other component, $v_{rg}$, is associated with the interface perturbation $\eta_r$:

\begin{equation}
    v_{rg} = \frac{D \eta_r}{Dt}.
\end{equation}

Through hydrostatic balance and geostrophic balance, the interface displacement $\eta_r$ is related to the streamfunction by \citep{yano1994jupiter,vallis2017atmospheric}

\begin{equation}
    g' \eta_r = 2\Omega ( \psi - \psi_{\mathrm{Ohmic}} ) \approx 2\Omega \psi,
\end{equation}

\noindent where $\psi_{\mathrm{Ohmic}}$ denotes the flow in the underlying Ohmic layer and we assume $|\psi_{\mathrm{Ohmic}}| \ll |\psi|$ as the motions in the Ohmic layer is very weak compared with the dynamically active layer. Therefore,

\begin{equation}
    v_{rg} = \frac{D \eta_r}{Dt} = \frac{2\Omega}{g'} \frac{D\psi}{Dt},
\end{equation}

Equation~\ref{eq:yanoPV} can thus be rewritten as

\begin{equation}\label{eq:final}
    \frac{D \nabla^2 \psi}{Dt} + \beta \frac{\partial \psi}{\partial x} = - \frac{2\Omega \tilde{\rho}(z_b)}{H_\rho} \sqrt{\frac{\nu_V}{4\Omega}} \nabla^2 \psi + \frac{4\Omega^2 \tilde{\rho}(z_b)}{g' H_\rho \sin{\phi}} \frac{D\psi}{Dt} = -\mu \nabla^2 \psi + \frac{1}{L_d^2} \frac{D \psi}{Dt},
\end{equation}

\noindent where $\mu = \tilde{\rho}(z_b) \sqrt{\nu_V \Omega}/H_\rho$ is an effective linear drag coefficient, representing the influence of Ohmic dissipation, and $L_d = \sqrt{g'H_\rho \sin{\phi}/(4\Omega^2\tilde{\rho}(z_b))}$ is the deformation radius. This formulation is analogous to that used in previous studies to represent coupling between different layers \citep[e.g.,][]{marcus1988numerical}, although in prior work, $L_d$ typically represents the coupling between the stratified weather layer and the deep convective layer, whereas here it represents the coupling between the entire dynamical layer and the underlying Ohmic layer.

The deformation radius $L_d$ depends on latitude. For simplicity, we assume a constant deformation radius $L_d$ to represent the effect of stratification and neglect its latitudinal dependence. We estimate $L_d$ based on the observed gravity wave speed $c_g$. On Jupiter, $c_g \approx 450~\mathrm{m}~\mathrm{s}^{-1}$ \citep{ingersoll1995waves}, implying $L_d = c_g/(2\Omega) \approx 1300~\mathrm{km}$. In our simulations, we set $L_d$ as a fixed fraction of the Rhines scale using Jupiter-like values as a reference. $\mu$ is a function of $H_\rho$, which also varies with latitude. For better control of the total kinetic energy, we set $\mu$ to a constant value so that the kinetic energy scales as $\varepsilon / \mu$, where $\varepsilon$ is the energy injection rate per unit mass.

The system is energized by convection driven by intrinsic heating and by baroclinic instability arising from differential solar forcing. Rather than resolving these convective and baroclinic instabilities explicitly, we represent their effects as stochastic forcing at a prescribed wavenumber \citep{smith2004local}:

\begin{equation}\label{eq:stochastic}
    \mathcal{F} = \sqrt{\frac{2\varepsilon n_e^2}{\delta t}} \Sigma_m \frac{ e^{-\frac{(n_m-n_e)^2}{2\delta n^2}} e^{i (n_x x+n_yy +\varphi_m)}}{C_\mathcal{F}},
\end{equation}

\noindent where $\mathcal{F}$ is the eddy forcing term, $n_e$ is the forcing wavenumber, $n_m = (n_x^2 + n_y^2)^{1/2}$ is the total wavenumber, $n_x$ and $n_y$ are wavenumbers in $x$ and $y$ directions, $\delta n$ is the forcing wavenumber bandwidth, $\delta t$ is the timestep, $\varphi_m$ is a random phase updated each timestep, and $C_\mathcal{F}$ normalizes the sum to unit amplitude. We apply a constant forcing amplitude across all latitudes for simplicity. To allow sufficient scale separation in the inverse energy cascade, we choose the forcing length scale to be 4 grid points, i.e. $L_e=4L_{grid}$. For numerical stability, we also introduce a hyperviscosity term to suppress grid-scale noise. The full governing equation becomes:

\begin{equation}\label{eq:introduceFD}
    \frac{D}{Dt} \left( \zeta - \frac{\psi}{L_d^2} \right) + \beta \frac{\partial \psi}{\partial x} = \mathcal{F} - \mu \zeta - \nu_h \nabla^4 \zeta,
\end{equation}

\noindent where $\zeta = \nabla^2 \psi$ is the vorticity of the flow and $\nu_h$ is the hyperviscosity coefficient.

The $\beta$ profile reflects both spherical geometry and density stratification, the latter of which remains poorly constrained by observations of giant planets, particularly Uranus and Neptune. While stronger stratification increases the magnitude of $\beta$, it does not alter its sign or qualitative shape (c.f. Fig.~5 in \cite{gastine2014zonal}), and is therefore not expected to influence the direction of the equatorial jet. For simplicity and numerical stability, we assume weak stratification and the $\beta$ effect is dominated by spherical geometry:

\begin{equation}\label{eq:beta}
\beta = - \frac{2\Omega}{H} \frac{\partial H}{\partial y} =
\begin{cases} 
- 2\Omega \cos{\phi}/(R\sin^2{\phi})& \mathrm{if}\ |\phi| < |\phi_c|, \\
2 \Omega \cos{\phi}/ (|\sin{\phi}| [(R-D)^2 - R^2 \cos^2{\phi}]^{1/2}) & \mathrm{if}\ |\phi| > |\phi_c|,
\end{cases}
\end{equation}

\noindent where

\begin{equation}\label{eq:H}
H =
\begin{cases} 
R |\sin{\phi}|  & \mathrm{if}\ |\phi| < |\phi_c|, \\
R |\sin{\phi}| - [ (R-D)^2 - R^2 \cos^2{\phi} ]^{1/2} & \mathrm{if}\ |\phi| > |\phi_c|,
\end{cases}
\end{equation}

\noindent $R$ is the planetary radius, $D$ is the jet penetration depth, and $\phi_c = \arccos((R-D)/R)$ is the latitude of the TC. This $\beta$ profile exhibits a singularity at the equator ($\phi = 0$) and a discontinuity at $\phi = \phi_c$, which can lead to numerical instability. To mitigate this issue, we exclude the equatorial region from 15$^\circ$S to 15$^\circ$N, where the rapid increase in $\beta$ leads to numerical instability. We also apply smoothing near the TC using a polynomial fit to ensure $H$ and $\beta$ are both continuous and differentiable. These modifications do not change the sign of $\beta$ or its derivative and thus are not expected to qualitatively affect the system behavior. The final $\beta$ profiles are shown in Fig.~\ref{fig:sketch-beta-h}.

Outside the TC, the atmosphere does not interact with the Ohmic dissipation layer, resulting in $\mu = 0$. However, reducing $\mu$ substantially increases computational time. The model's equilibrium timescale is proportional to $1/\mu$. The jet speed can be estimated as $(\varepsilon/\mu)^{1/2}$, allowing the maximum time step to be estimated as $\delta t = k_{CFL} L_{grid} \varepsilon^{-1/2} \mu^{1/2}$ according to the Courant–Friedrichs–Lewy (CFL) condition, where $k_{CFL}=0.2$ is a coefficient chosen to ensure numerical stability. Consequently, the total number of time steps can be estimated as $\varepsilon^{1/2}/(k_{CFL} L_{grid} \mu^{3/2})$. As a result, small values of $\mu$ significantly increase computational time and make it difficult for equatorial jets to reach equilibrium. To balance realism and computational efficiency, we reduce the damping coefficient outside the TC to 10\% of its value inside. The hyperviscosity coefficient is scaled using dimensional analysis as $\nu_h = C_\nu \varepsilon^{1/3} L_{\text{grid}}^{10/3}$, where $C_\nu$ is a tuning parameter adjusted to ensure numerical stability in each simulation.

Equilibrating simulations with low energy injection rates ($\varepsilon$) and damping coefficients ($\mu$) is computationally challenging. To mitigate this, we initially run simulations with both $\varepsilon$ and $\mu$ increased by a factor of 1000, to accelerate the approach to equilibrium. In these high-energy-injection and high-damping simulations, we observe some poleward migration of the mid-latitude jets \citep{chemke2015poleward}, but the direction and magnitude of the equatorial jet remain unchanged in the final state (see Fig.~\ref{figS2:IC} for examples of jet migration). These simulations typically require $10^5$–$10^7$ rotation periods to reach a quasi-equilibrium state, where the total kinetic energy remains unchanged and the total energy dissipation rate matches the energy injection rate. Once this state is reached, we continue the simulations using the smaller, target values of $\varepsilon$ and $\mu$. Despite this approach, the flow adjustment remains computationally expensive, so each simulation is run until the zonal-mean zonal velocity shows no significant long-term changes. In the low-energy-injection and low-damping cases, the jets remain steady without noticeable migration.

For simulations with a finite deformation radius $L_d$, the system must maintain a ratio between kinetic and potential energy \citep{vallis2017atmospheric}. In these cases, while the kinetic energy typically reaches a quasi-equilibrium state relatively quickly, the potential energy equilibrates on a much longer timescale, possibly due to the absence of explicit damping in the potential energy component of our model. As a result, due to computational resource constraints, the potential energy in some simulations may not fully equilibrate, while we ensure that the kinetic energy in all simulations reaches quasi-equilibrium.

\begin{figure*}
\centering
\includegraphics[width=0.73\textwidth]{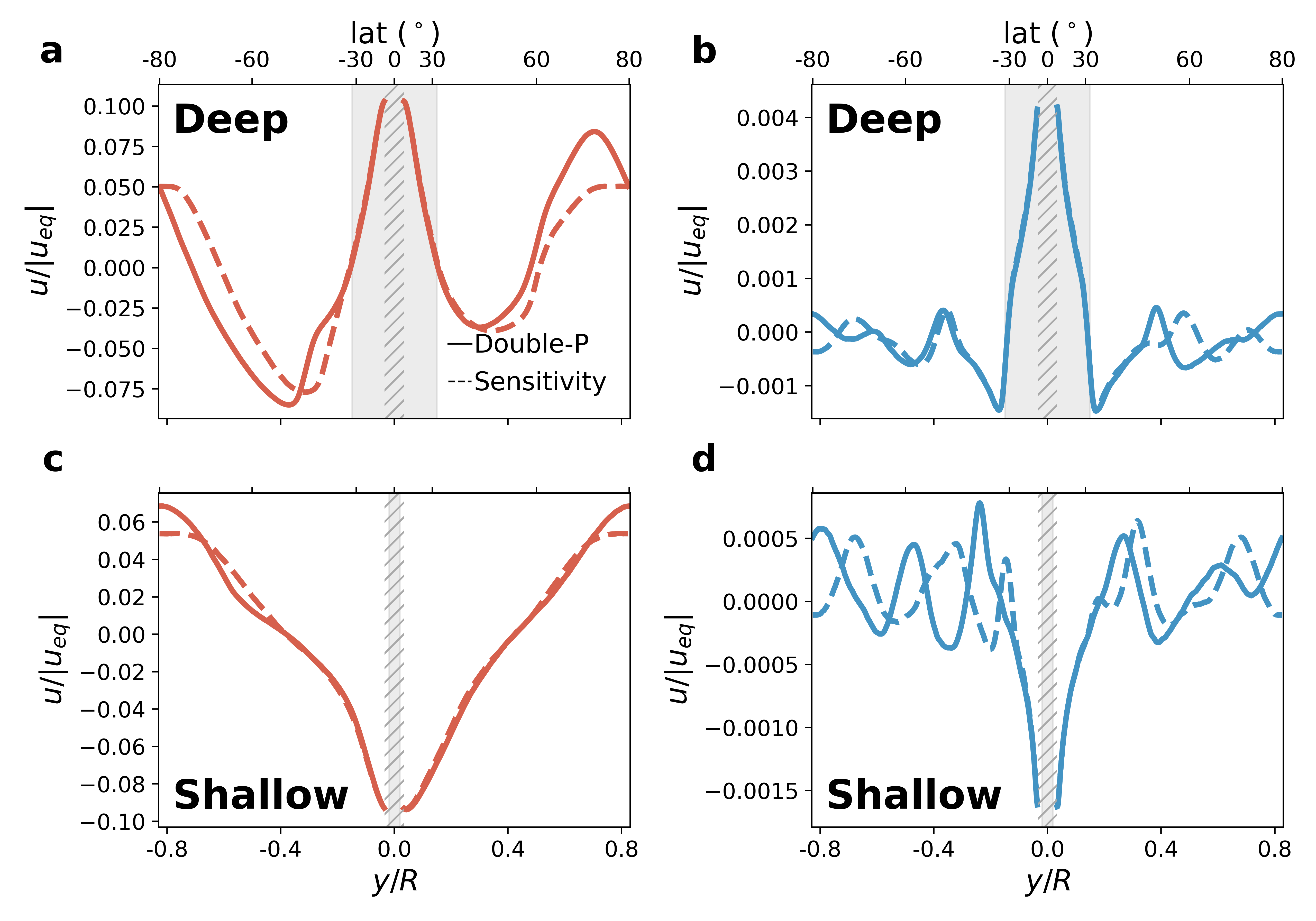}
\caption{\textbf{Sensitivity to boundary conditions.} Zonal-mean zonal velocity fields from simulations with doubly periodic boundary conditions (solid, ``Double-P'') and with sponge layers that damp vorticity to zero at the northern and southern boundaries (dashed, ``Sensitivity''). Results are shown for simulations with $L_d \to \infty$, using $\varepsilon = 5 \times 10^{-7}$ and $\mu = 5 \times 10^{-5}$ in \textbf{a} and \textbf{c}, and $\varepsilon = 10^{-10}$ and $\mu = 10^{-4}$ in \textbf{b} and \textbf{d}.}
\label{figS:boundary_condition}
\end{figure*}

We applied doubly periodic boundary conditions for numerical efficiency, which connect the southernmost and northernmost points of the domain and are hence not physically realistic. To assess the impact of this boundary condition, we performed sensitivity tests that introduced a sponge layer near the northern and southern boundaries, where the vorticity is damped to zero, thereby ensuring separation between the two hemispheres at the highest latitudes. We tested this configuration using simulations with an infinite deformation radius ($L_d \to \infty$) and parameter pairs ($\varepsilon=5\times10^{-7}$, $\mu=5\times10^{-5}$) and ($\varepsilon=10^{-10}$, $\mu=10^{-4}$). The resulting zonal-mean zonal velocity fields differ only in the positions of the mid-latitude jets, while the equatorial jet direction, the overall jet strength, and the number of jets remain unchanged (Fig~\ref{figS:boundary_condition}), confirming that the chosen boundary conditions do not affect the results presented here.

\section{Energy injection rate and length scales}\label{app:epsilon}

For each of the four giant planets, we estimate their energy injection rate ($\varepsilon$) and thereby their eddy length scale ($L_e$), $\beta$ scale ($L_\beta$), and Rhines scale at the equator ($L_{Rh,eq}$) and mid-latitudes ($L_{Rh}$), as listed in Table~\ref{tab:planet_parameter}. The major energy sources for these systems include intrinsic heating and solar heating. Estimating the contribution from differential solar forcing is challenging, however, because it depends on the excited vertical mode, which remains poorly constrained. To obtain an order-of-magnitude estimate of the energy injection rate for the four giant planets, we therefore calculate it based on the intrinsic heat flux. Treating a giant planet’s atmosphere as an anelastic fluid, the globally integrated energy equation is given by \citep{vallis2017atmospheric}:

\begin{equation}\label{eq:energy_anelastic}
    \frac{\partial}{\partial t} \int \frac{1}{2} \tilde{\rho} \boldsymbol{v}^2 dV = \int \tilde{\rho} v_r b_a dV - E_{dis},
\end{equation}

\noindent where $\tilde{\rho} = \tilde{\rho}(r)$ is the reference density stratification, $b_a=g \theta/\theta_0$ represents the entropy anomaly, $\theta$ is potential temperature anomaly, and $\theta_0$ is a reference potential temperature. In equilibrium, the left-hand side is zero, meaning that the first term on the right-hand side represents the energy injection rate, which balances the dissipation rate $E_{dis}$. The entropy anomaly equation gives

\begin{equation}\label{eq:entropy}
    \frac{D b_a}{D t} = \frac{g}{\theta_0} \kappa \nabla^2 \theta,
\end{equation}

\noindent where $\kappa$ is thermal diffusivity. Multiplying Equation~\ref{eq:entropy} by $\tilde{\rho} r$ and integrating it over the domain, and assuming entropy conservation $\int [\tilde{\rho} D(b_a r) / Dt] dV = 0$, gives

\begin{equation}\label{eq:theta_integral}
\begin{aligned}
\int \tilde{\rho} v_r b_a dV &= -\frac{g}{\theta_0} \int \tilde{\rho} r \kappa \nabla^2 \theta dV \\
&= -\frac{g}{\theta_0} \int [\kappa \nabla \cdot (\nabla \theta \tilde{\rho} r) - \kappa \nabla \theta \cdot \nabla(\tilde{\rho} r)] dV \\
&= \frac{g}{c_p} \left( \frac{Q_{t} r_{t} A_{t}}{T_{t}} -  \frac{Q_{b} r_{b} A_{b}}{T_{b}} \right) + \frac{g}{\theta_0} \int \kappa \frac{\partial \theta}{\partial r} \frac{\partial (\tilde{\rho} r)}{\partial r} dV,
\end{aligned}
\end{equation}

\noindent where we used Gauss's theorem and the boundary condition that $\kappa \nabla \theta = -(Q \theta_0)/(\tilde{\rho} c_p T)$. Here, $c_p$ is the heat capacity, $Q$ the heat flux, $A$ the area, $T$ the temperature, and the subscripts $t$ and $b$ refer to the top and bottom boundaries, respectively. We neglect the contribution by thermal diffusion (second term in Equation~\ref{eq:theta_integral}) and disregard the area differences between the surface and the bottom, i.e., $A_{t}\approx A_{b}= A$. We assume the entropy flux $Q/T$ is a constant, i.e., $Q_{b}/T_{b}=Q_{t}/T_{t}$, the contribution from boundary fluxes can be expressed by $ (g Q_{t} D A)/(c_p T_{t})$, where $D$ is the penetration depth of the jet. The energy injection rate per unit mass can then be written as
\begin{equation}\label{eq:varepsilon_equation}
    \varepsilon = \frac{\int \tilde{\rho} v_r b_a dV}{M_{atm}} = \frac{Q_{t} g}{\langle \tilde{\rho} \rangle c_p T_{t}}, \ \ \ \langle \tilde{\rho} \rangle = \frac{4 \pi \int_{R-D}^{R} \tilde{\rho} r^2 dr}{D A},
\end{equation}

\noindent where $M_{atm}$ is the total atmosphere mass, $R$ is the planetary radius, and $\langle \tilde{\rho} \rangle$ is the averaged density. Assuming an ideal gas atmosphere with a dry adiabatic lapse rate, we calculate the energy injection rate per unit mass as follows (for simplicity, we omit the spherical metric):


\begin{equation}\label{eq:varepsilon}
    \varepsilon = \frac{Q_{t} g^2 D}{c_p T_{t} p_{t}\left[ \left( 1 + \frac{gD}{c_p T_{t}} \right)^\gamma - 1 \right]},
\end{equation}

\noindent where $p_{t}$ is the pressure at the upper boundary, $\gamma=c_p/R_d=3.5$ is the adiabatic index, and $R_d$ is the gas constant. We choose $p_{t}=0.1$~bar and $T_{t}$ as the blackbody equivalent temperature set by solar radiation, and the heat flux $Q_{t}$ is set to the measured intrinsic heat flux near the surface. We calculate the energy injection rate in all four planets using the planetary parameters listed in Table~\ref{tab:planet_parameter}.

Using the energy injection rate $\varepsilon$, planetary rotation rate $\Omega$, and jet penetration depth $D$, we estimate the influence of rotation on convection. Following \cite{aurnou2020connections}, the convective Rossby number $\mathrm{Ro_C}$ serves as a useful indicator of rotational constraint. In the rapidly rotating regime, where $\mathrm{Ro_C} \ll 1$, convection is strongly influenced by rotation, leading to columnar structures aligned with the rotation axis. In contrast, in the non-rotating or slowly rotating regimes, convection is more isotropic. The convective Rossby number can be estimated using the modified flux Rayleigh number, defined as $\mathrm{Ra_F^*} = \varepsilon (2\Omega)^{-3} D^{-2}$. Depending on the rotational regime, the scaling differs: $\mathrm{Ro_C} \sim \mathrm{Ra_F^*}^{1/3}$ in the slowly rotating limit, and $\mathrm{Ro_C} \sim \mathrm{Ra_F^*}^{1/5}$ in the rapidly rotating limit. Applying planetary parameters for the four giant planets (Table~\ref{tab:planet_parameter}), we find that $\mathrm{Ra_F^*} \ll 1$ in all cases, implying $\mathrm{Ro_C} \ll 1$ and confirming that convection occurs in the rapidly rotating limit. Accordingly, we compute $\mathrm{Ro_C} \sim \mathrm{Ra_F^*}^{1/5}$, with results listed in Table~\ref{tab:planet_parameter}.

The eddy length scale $L_e$ is determined based on rotating convection scales, given by $L_e = \varepsilon^{1/4} D^{1/2} (2\Omega)^{-3/4}$ \citep{jones1993convection}. The $\beta$ scale $L_\beta$ and the frictional Rhines scale $L_{Rh}$ are calculated following \cite{galperin2008zonostrophic}: $L_\beta = 2 \pi \varepsilon^{1/5} \beta^{-3/5}$ and $L_{Rh} = \pi (2U/\beta)^{1/2}$, where $U$ is chosen as the characteristic flow speed at mid-latitudes on the four planets. Since $\beta$ varies by a factor of 100 across the domain, we calculate the averaged $L_\beta$ and $L_{Rh}$ over latitudes from 5$^\circ$ to 85$^\circ$. For the equatorial Rhines scale $L_{Rh,eq}$, we use the typical equatorial jet speed and the $\beta$ value at 15$^\circ$.

\section{Calculation of the eddy momentum flux using SSST}\label{app:S3T}

We decompose the flow into zonal-mean jets $U$ and eddies described by the perturbation streamfunction $\psi'$, written as a Fourier expansion in zonal harmonics,

\begin{equation}
    \psi' (x,y,t) = \sum_{n_x} \psi_n(y,t) e^{i n_x x}.
\end{equation}

Following SSST \citep{farrell2003structural}, the evolution of the ensemble-averaged covariance matrix of the perturbation field, $\boldsymbol{C}\equiv<\psi \psi^\dagger>$, together with the mean flow, satisfies

\begin{equation}\label{eq:C_S3T}
    \frac{d \boldsymbol{C}}{dt} = \boldsymbol{A} \boldsymbol{C} + \boldsymbol{C} \boldsymbol{A}^\dagger + \boldsymbol{Q}, 
\end{equation}

\begin{equation}\label{eq:U_S3T}
    \frac{d U}{dt} = - \frac{{n_x}}{2} \mathrm{diag[\mathrm{Im(\boldsymbol{C} \boldsymbol{\Delta}^\dagger})]} - \mu \boldsymbol{U} - \nu_h  \boldsymbol{D^2 U}, 
\end{equation}

\noindent where 

\begin{equation}
    \boldsymbol{A} (\boldsymbol{U}) = - i n_x [\boldsymbol{U} + [\beta \boldsymbol{I} - \boldsymbol{D}^2 \boldsymbol{U}] \boldsymbol{\Delta}^{-1}] - \mu \boldsymbol{I} - \nu_h \boldsymbol{\Delta}^2,
\end{equation}

\noindent $\boldsymbol{I}$ is the identity matrix, $\boldsymbol{D}^2$ represents $d^2/dy^2$, $\boldsymbol{\Delta}\equiv \boldsymbol{D}^2 - n_x^2 \boldsymbol{I}$, $\boldsymbol{Q}$ is the stochastic forcing covariance, and $\dagger$ denotes Hermitian transpose. As shown in Equation~6 in \cite{farrell2003structural}, the zonal-mean eddy momentum flux convergence is given by

\begin{equation}
    \overline{vq} = - \sum_{n_x} \frac{{n_x}}{2} \mathrm{diag[\mathrm{Im(\boldsymbol{C} \boldsymbol{\Delta}^\dagger)}]},
\end{equation}

\noindent which is the first term on the right-hand-side of Equation~\ref{eq:U_S3T} that acts to accelerate the jets.

To isolate the influence of the planetary $\beta$ profile on the eddy momentum flux convergence, we diagnose the statistically stationary covariance by setting the left-hand-side of Equation~\ref{eq:C_S3T} to zero and evaluating the system about the state $U=0$ to solve for $\boldsymbol{C}$. This calculation is intended to quantify the linear statistical tendency imposed by the prescribed $\beta$ profile (i.e., the sign and structure of $\overline{vq}$) for the initial emergence of the equatorial jet in the SSST framework, rather than to reconstruct the fully equilibrated jet structure obtained in the nonlinear simulations. We include zonal wavenumbers from $n_x=n_0$ to $n_x=32n_0$, where $n_0=2\pi/L_x$. The resulting $\overline{vq}$ profiles are shown in Figs.~\ref{fig:Mechanism_compare}a-b and \ref{fig:S3T}c-d.


\section{Diagnostic of the jet width and Rhines scale}\label{app:results}

\begin{figure*}
\centering
\includegraphics[width=0.88\textwidth]{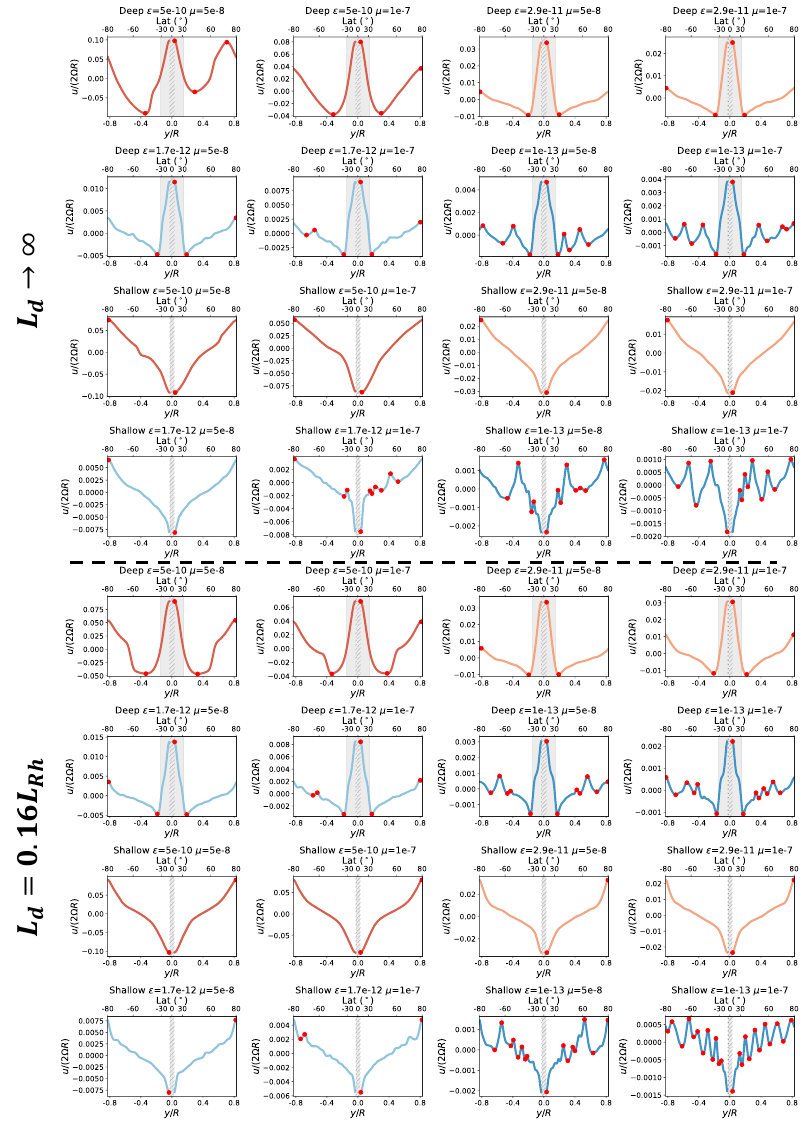}
\caption{\textbf{Zonal-mean jet profiles.} Lines represent the zonal-mean zonal velocity $\overline{u}$ from the simulations. Red dots indicate the diagnosed jet peaks, as defined in APPENDIX~\ref{app:results}. Gray shading indicates regions outside the TC, and the gray hatched regions mark the area excluded from our simulation domain. Note that the lowest latitudes in the northern and southern hemispheres ($\pm15^\circ$) are connected as a single point in our simulation.}
\label{figS1:U}
\end{figure*}

\begin{figure*}
\centering
\includegraphics[width=0.95\textwidth]{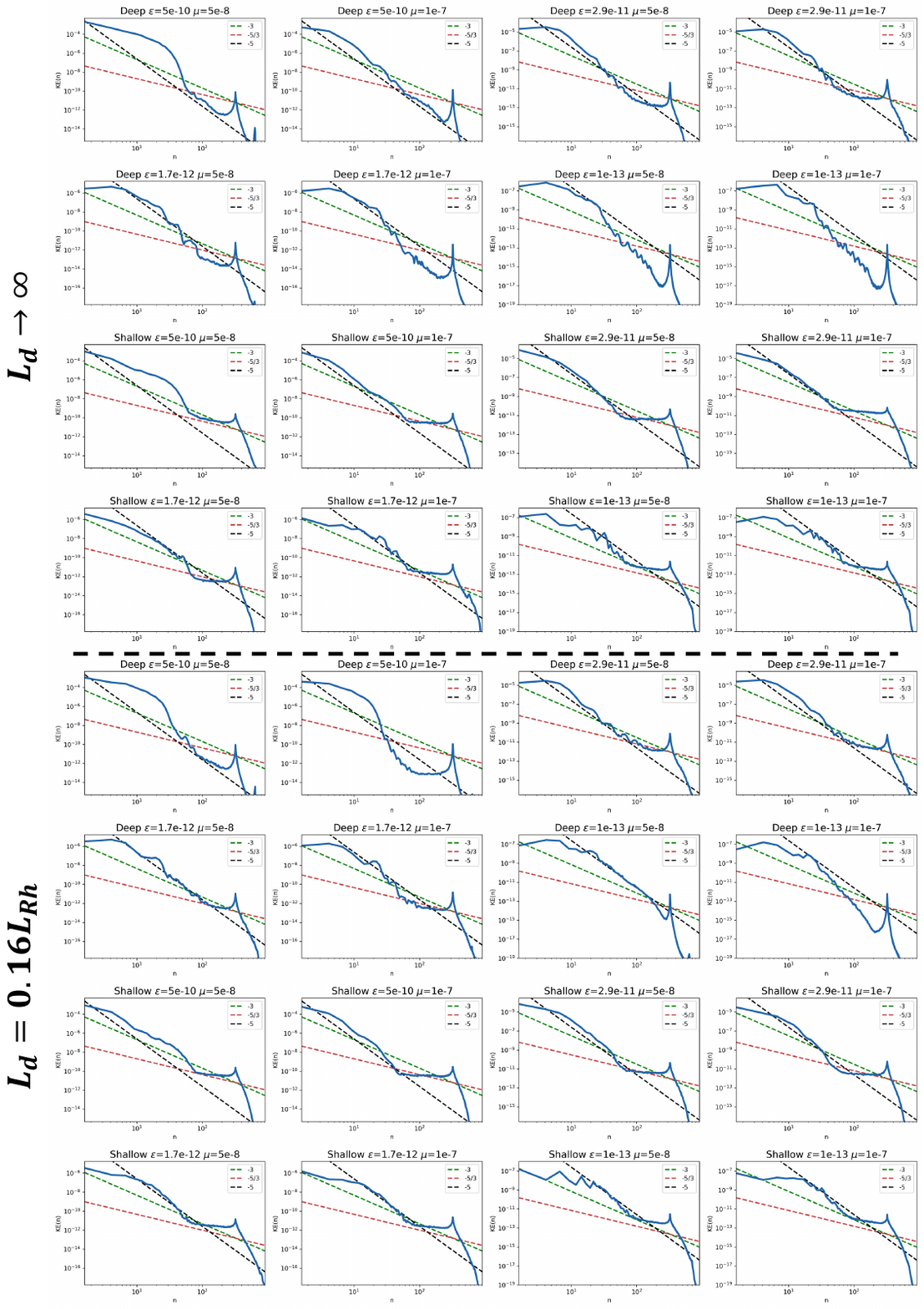}
\caption{\textbf{Kinetic energy spectra.} Snapshots of the kinetic energy (KE; $u^2/2 + v^2/2$) spectra in all simulations. The blue line shows the simulated KE spectrum. The green, red, and black dashed lines indicate the theoretical slopes of $-3$, $-5/3$, and $-5$, respectively. The transition wavenumbers $n_e$ and $n_\beta$ are indicated by the intersections of the green and red lines, and the red and black lines, respectively.}
\label{figS:Allspectrum}
\end{figure*}

We diagnose jets as local maxima and minima in the zonal-mean, time-averaged zonal velocity profile $\overline{u}$, excluding extrema with a velocity difference smaller than 3\% of the maximum difference (Fig.~\ref{figS1:U}). Jet widths are defined as the distances between the midpoints of adjacent extrema. To compute the modified Rhines scale, we calculate $\beta$ as the average value within the jet region. The velocity scale $U$ in the Rhines scale is formally defined as the root-mean-square velocity \citep{rhines1975waves}. Here, we define $U$ as the velocity difference between the jet core and its boundaries, since most of the energy is concentrated in the jets, and the jet speed is more directly observable on the giant planets. For reference, the kinetic energy spectra (Fig.~\ref{figS:Allspectrum}) in all simulations are also provided here.

\newpage

\bibliography{Jet}{}
\bibliographystyle{aasjournalv7}

\end{document}